\documentclass[12pt]{article}
\usepackage{amsmath,amssymb,epsfig}
\newcommand{\neu}[1]{\tilde{\chi}^0_{#1}}
\newcommand{\cha}{\tilde{\chi}}
\newcommand{\st}{\tilde{t}}
\newcommand{\mneu}[1]{m_{\tilde{\chi}^0_{#1}}}
\newcommand{\mcha}[1]{m_{\tilde{\chi}^\pm_{#1}}}
\newcommand{\sbot}{\tilde{b}}

\begin{document}

\begin{titlepage}
\pagestyle{empty} \baselineskip=21pt
\renewcommand{\thefootnote}{\fnsymbol{footnote}}

\rightline{IPPP/09/55} \rightline{DCPT/09/110} \vskip 0.5in
\begin{center}
{\large{\bf Towards Measuring the Stop Mixing Angle at the LHC}}
\end{center}
\begin{center}
\vskip 0.2in {\bf Krzysztof Rolbiecki, Jamie Tattersall} and {\bf
Gudrid Moortgat-Pick} \vskip 0.1in

{\it
{Institute for Particle Physics Phenomenology, \\
Durham University, Durham DH1 3LE, UK}}
 \vskip 0.5in

\vskip 0.5in {\bf Abstract}
\end{center}
\baselineskip=18pt \noindent
We address the question of how to determine the stop mixing angle
and its CP-violating phase at the LHC. As an observable we discuss
ratios of branching ratios for different decay modes of the light
stop $\tilde{t}_1$ to charginos and neutralinos. These observables
can have a very strong dependence on the parameters of the stop
sector. We discuss in detail the origin of these effects. Using
various combinations of the ratios of branching ratios we argue
that, depending on the scenario, the observable may be promising in
exposing the light stop mass, the mixing angle and the CP phase.
This will, however, require a good knowledge of the supersymmetric
spectrum, which is likely to be achievable only in combination with
results from a linear collider.
\bigskip \\ \\

\vfill
\begin{flushleft}\baselineskip=15pt
e-mail: \texttt{\\krzysztof.rolbiecki@desy.de \\
jamie.tattersall@th.physik.uni-bonn.de \\
gudrid.moortgat-pick@desy.de}
\end{flushleft}
\end{titlepage}
\setcounter{page}{2} \baselineskip=18pt

\section{Introduction}

Supersymmetry (SUSY)~\cite{susy} is one of the best-motivated
extensions of the Standard Model (SM). It allows one to stabilize
the hierarchy between the electroweak (EW) scale and the Planck
scale and to naturally explain electroweak symmetry breaking (EWSB)
by a radiative mechanism. The naturalness of the scale of
electroweak symmetry breaking and the Higgs mass places a rough
upper bound on the superpartner masses of several TeV and the fits
to the electroweak precision data point to a rather light SUSY
spectrum~\cite{Ellis:2007fu}.  Therefore a high potential for the
discovery of SUSY is expected at the
LHC~\cite{Ball:2007zza,Aad:2009wy,Hinchliffe:1996iu}.

Strongly interacting SUSY particles will be produced copiously at
the LHC, with cross sections up to tens of pb for squarks and
gluinos if their mass range is of a few hundreds GeV. Cross sections
for direct production of scalar top quarks -- the supersymmetric
partners of top quarks -- are expected to be smaller due to a
different production mechanism, however, still in a range of a few
pb, e.g.\ for the SPS1a$'$ parameter
point~\cite{AguilarSaavedra:2005pw}. The other possible source of
3rd generation squarks, depending on the details of particle mass
spectrum in the SUSY scenario, would be decays of other squarks and
gluinos~\cite{Hisano:2003qu}.

Stops are of a special interest since they play an important role in
the mechanism of radiative electroweak symmetry breaking. Light
stops together with CP-violating phases can also provide an
attractive mechanism for electroweak baryogenesis by triggering a
strong first-order electroweak phase
transition~\cite{Carena:2008vj}. Therefore a careful analysis of the
stop sector can give an insight into the mechanism of EWSB and the
origin of the matter-antimatter asymmetry. Finally, the stop sector
has a large impact on the masses of the Higgs
bosons~\cite{Ellis:1990nz}, and in the presence of the CP-violating
phases it triggers mixing between CP-odd and CP-even Higgs
states~\cite{Pilaftsis:1998pe}. Therefore a precise knowledge of the
3rd generation squark parameters would allow us to test the anatomy
of the Higgs boson sector and electroweak symmetry breaking in the
MSSM.

If stops are within the kinematic reach of the International Linear
Collider or CLIC, production cross sections can be measured with a
high accuracy~\cite{Bartl:2000kw,Freitas:2007zr,Kovarik:2004vd}.
Using polarized beams, this can provide information on the mixing
angle and masses, and a precise determination of stop sector
parameters can be foreseen~\cite{Bartl:1996wt,kramlphd,finch}. In
this paper, however, we concentrate on the measurement of the stop
sector at the hadron collider. One of the advantages of hadron
colliders, like the LHC, is the enhancement of cross section due to
strong interactions~\cite{Baer:1991cb,beenakker98}. On the other
hand, however, further challenges will become important due to the
harsh experimental environment.

One of the possible sources of stops at the LHC will be decays of
other supersymmetric particles, for example gluinos. The analysis of
kinematical edges in the invariant mass distributions of the cascade
decay chains provides an example of one measurement that is often
studied at the
LHC~\cite{Hisano:2003qu,Weiglein:2004hn,Casadei:2010nf}. Taking the
SPS1a$'$ scenario as an example, a large number of stops and
sbottoms will appear in the gluino decay chain. Both, however, can
give a similar experimental signature and consequently one has to do
a simultaneous analysis of sbottom and stop sectors. This leads to
good constraints for the sbottom sector but the constraints on the
stop mixing angle are much weaker. Another possible observable is
the polarization of top quarks in the decay $\st_1 \to \neu{1}\,t$.
The information on the stop mixing angle can be extracted here from
the forward-backward asymmetries in leptonic and hadronic top
decays~\cite{Perelstein:2008zt}. The feasibility of this method
depends on sufficient suppression of $t\bar{t}$ background that
might turn out to be difficult, see however~\cite{Meade:2006dw}.

Another way of getting access to stop parameters are global fits
using low energy and collider
data~\cite{Bechtle:2009ty,Lafaye:2007vs}. This method gives very
good results when the fit is done within highly constrained models
like mSUGRA, see e.g.\ Fittino~\cite{Bechtle:2009ty}. In this
analysis the input observable was an invariant-mass end-point in the
already mentioned gluino decay chain. However, when analyzing a MSSM
model with 18 free parameters the constraints for the third
generation squarks mass parameters and the trilinear coupling
parameters are rather poor. Therefore, one has to study whether
adding a new observable would allow us to achieve better constraints
from such fits. In the general MSSM there are 105 free parameters
and in principle all of these should be determined separately to
fully realize the model of SUSY breaking. Therefore, it is very
important to perform a model independent measurement of the stop
mixing angle.

In this paper we focus our attention on the decays of the light top
squark to charginos and neutralinos that are possible in a wide
range of scenarios of the Minimal Supersymmetric Standard Model
(MSSM):
\begin{eqnarray}
&& \tilde{t}_1 \to \tilde{\chi}_i^+\, b \,, \label{eq:chab}\\
&& \tilde{t}_1 \to \tilde{\chi}^0_j\, t \,. \label{eq:neu1t}
\end{eqnarray}
The stop and sbottom decays have already been analyzed in the
literature in some
detail~\cite{kramlphd,Bartl:1994bu,Hidaka:2000cm,Bartl:2003pd},
including radiative
corrections~\cite{nlo1,nlo2,nlo3,Bartl:1997pb,Guasch:2001kz,full1loop,nlo4}.
In this paper we propose a method to expose the properties of the
stop sector using simultaneously the decays Eqs.~\eqref{eq:chab}
and~\eqref{eq:neu1t}. Observation of the above channels will be
experimentally challenging. However recent studies suggest it may be
feasible in the case of $\tilde{t}_1 \to \tilde{\chi}^0_1\, t $
followed by hadronic top decays \cite{Meade:2006dw,Plehn:2010st}, as
well as $\tilde{t}_1 \to \tilde{\chi}^0_2\, t $
\cite{MoortgatPick:2010wp}. Certainly, more detailed studies are
required, but the new technique of top-tagging~\cite{Plehn:2009rk}
can significantly improve the sensitivity.

We analyze three scenarios of the MSSM with different
gaugino/higgsino composition of charginos and neutralinos and
discuss in detail the relevance of the underlying gaugino/higgsino
mixing for the determination of the stop mixing angle. We show that
the branching ratios for these decays can be a sensitive probe of
the mixing angle in the stop sector and also of the CP-violating
phase. The highest sensitivity can be obtained in scenarios with
weak mixing between gauginos and higgsinos, for instance in mSUGRA,
as explained later in detail. We use a model-independent approach,
i.e.\ without assuming a particular structure for the stop mass
matrix, and parametrize the stop interactions in terms of the mixing
parameters $\cos\theta_{\st}$ and $\phi_{\st}$. Since the absolute
measurement of branching ratios is expected to be very difficult at
the LHC we propose to exploit another set of observables, namely
ratios of branching ratios, cf.\
Ref.~\cite{Hisano:2003qu,Weiglein:2004hn}. We argue that if more
than one of the above decay modes can be observed at the LHC, it may
be possible to derive some constraints on the mass and the mixing
parameters of stops. We briefly discuss possible experimental issues
for these processes. Finally, a $\chi^2$ fit is performed to give a
range for the expected parameter determination precision under the
assumption of a clean signal sample.

In order to successfully extract the parameters of the stop sector,
one is going to need certain information about the structure of the
chargino and neutralino sectors. In some scenarios this task may be
difficult to achieve at the LHC due to its limited precision. In
such a case, the input from a future linear collider will be needed,
providing a measurement of the masses and mixing angles of gauginos
and higgsinos~\cite{Weiglein:2004hn,Desch:2003vw}.

The paper is organized as follows. In Section~\ref{sec:mixing} we
give a brief overview of the mixing and the couplings of the stop,
chargino and neutralino sectors of the MSSM. In
Section~\ref{sec:decays} we give analytic expressions for the decay
widths of the light stop into charginos and neutralinos and analyze
their dependence on the stop mixing parameters in chosen scenarios.
Section~\ref{sec:results} explains in detail how to determine the
stop mixing parameters using stop decays at the LHC for our
benchmark models. Finally, we summarize our findings in
Section~\ref{sec:conclusions}.

\section{Sparticle mixing and couplings}\label{sec:mixing}
\subsection{Stop sector of the MSSM}

In the Minimal Supersymmetric Standard Model the stop sector is
defined by the mass matrix ${\cal M}_{\tilde{t}}$ in the basis of
gauge eigenstates $(\tilde{t}_L, \tilde{t}_R)$. The $2 \times 2$
mass matrix depends on the soft scalar masses $\widetilde{M}_Q$ and
$\widetilde{M}_U$, the supersymmetric higgsino mass parameter $\mu$,
and the soft SUSY-breaking trilinear coupling $A_t$. It is given as
\begin{eqnarray}
    {\cal M}^2_{\tilde{t}}=
     \left( \begin{array}{cc}
                m_t^2 + m_{LL}^2 & {m_{LR}^*}\, m_t \\
                {m_{LR}}\, m_t     & m_t^2 + m_{RR}^2
            \end{array} \right)\, ,
\end{eqnarray}
where
\begin{eqnarray}
  m_{LL}^2 &=& \widetilde{M}_{Q}^2
    + m_Z^2\cos 2\beta\,( \frac{1}{2} - \frac{2}{3} \sin^2\theta_W )\, , \\
  m_{RR}^2 & =&  \widetilde{M}_{U}^2
                  + \frac{2}{3} m_Z^2 \cos 2\beta\, \sin^2\theta_W \, , \\
  m_{LR}    &= & {A_t} - {\mu^*} \cot\beta \,  ,
\end{eqnarray}
and $\tan\beta=v_2/v_1$ is the ratio of the vacuum expectation
values of the two neutral Higgs fields which break the electroweak
symmetry. From the above parameters only $\mu$ and $A_t$ can take
complex values in our convention\footnote{We work in a convention
where $M_2$ is real and positive. In the scalar stop sector only the
combination $\phi_\mu + \phi_{A_t}$ is a physical quantity, whereas
in the chargino/neutralino sector only $\phi_\mu$ and $\phi_1$
enter. Therefore we use $\phi_{A_t}$ and $\phi_\mu$ as independent
quantities for studying stop decays to charginos and neutralinos,
see e.g.\ Eqs.~\eqref{eq:phase}, \eqref{eq:massmatrix} and
\eqref{eq:mass_matrix}.}
\begin{eqnarray}\label{eq:at-mu}
A_t = |A_t|\, {\rm e}^{{\rm i}\phi_{A_t}},\qquad \mu = |\mu|\, {\rm
e}^{{\rm i}\phi_\mu}, \qquad (0\leq \phi_{A_t}, \phi_\mu < 2\pi)\,,
\end{eqnarray}
thus yielding CP violation in the stop sector.

The hermitian matrix ${\cal M}^2_{\tilde{t}}$ is diagonalized by a
unitary matrix $\mathcal{R}_{\tilde{t}}$ that rotates gauge
eigenstates, $\tilde{t}_L$ and $\tilde{t}_R$, into the mass
eigenstates $\tilde{t}_1$ and $\tilde{t}_2$:
\begin{eqnarray}
\mathcal{R}_{\tilde{t}}\: {\cal M}^2_{\tilde{t}}\:
\mathcal{R}_{\tilde{t}}^\dag = \left( \begin{array}{c c}
m^2_{\tilde{t}_1} & 0 \\ 0 & m^2_{\tilde{t}_2} \end{array}
 \right) \, ,
\end{eqnarray}
where we choose the convention $m^2_{\tilde{t}_1} <
m^2_{\tilde{t}_2} $ for the masses of $\tilde{t}_1$ and
$\tilde{t}_2$. The matrix $\mathcal{R}_{\tilde{t}}$ rotates the
gauge eigenstates, $\tilde{t}_L$ and $\tilde{t}_R$, into the mass
eigenstates $\tilde{t}_1$ and $\tilde{t}_2$ as follows
\begin{eqnarray} \label{eq:stop-mix}
\left( \begin{array}{c}
                {\tilde{t}}_1 \\
                {\tilde{t}}_2
\end{array} \right) = \mathcal{R}_{\tilde{t}}
\left( \begin{array}{c}
{\tilde{t}}_L \\
{\tilde{t}}_R
\end{array} \right) =
\left( \begin{array}{cc}
 \cos\theta_{\tilde{t}}  &
 \sin\theta_{\tilde{t}}\, \mathrm{e}^{-\mathrm{i}\phi_{\tilde{t}}}   \\
- \sin\theta_{\tilde{t}}\, \mathrm{e}^{\mathrm{i}\phi_{\tilde{t}}}&
 \cos\theta_{\tilde{t}}
            \end{array} \right)
\left( \begin{array}{c}
                {\tilde{t}}_L \\
                {\tilde{t}}_R
\end{array} \right) \, ,
\end{eqnarray}
where $\theta_{\tilde{t}}$ and $\phi_{\tilde{t}}$ are the mixing
angle and the CP-violating phase of the stop sector, respectively.
The masses are given by
\begin{eqnarray}
m_{\tilde{t}_{1,2}}^2 = \frac{1}{2} \left( 2 m_t^2 + m_{LL}^2 +
m_{RR}^2 \mp \sqrt{(m_{LL}^2 - m_{RR}^2)^2 + 4 |m_{LR}|^2 m_t^2}
\right) \,,
\end{eqnarray}
whereas for the mixing angle and the CP phase we have
\begin{eqnarray}
\cos\theta_{\tilde{t}} &=& \frac{-m_t|m_{LR}|}{\sqrt{m_t^2 |m_{LR}|^2+(m_{\tilde{t}_1}^2-m_{LL}^2)^2}}\,,\\
\sin\theta_{\tilde{t}} &=& \frac{m_{LL}^2-m_{\tilde{t}_1}^2}{\sqrt{m_t^2 |m_{LR}|^2+(m_{\tilde{t}_1}^2-m_{LL}^2)^2}}\,,\\
\phi_{\tilde{t}} &=& \arg (A_t - \mu^* \cot\beta)
\label{eq:phase}\,.
\end{eqnarray}
By convention we take $0\leq \theta_{\tilde{t}}<\pi$ and $0\leq
\phi_{\tilde{t}} < 2 \pi$. It must be noted that $\phi_{\tilde{t}}$
is an \textquoteleft effective' phase and does not directly
correspond to the phase of any MSSM parameter. Instead, the phase
will have contributions from both $\phi_{A_t}$ and $\phi_{\mu}$ in
our particular convention. However, for $A_t \gg \mu \cot\beta$ one
has $ \phi_{\tilde{t}} \approx \phi_{A_t}$. If $m_{LL} < m_{RR}$
then $\cos^2\theta_{\tilde{t}}
> \frac{1}{{2}}$ and $\tilde{t}_1$ has a predominantly left gauge
character. On the other hand, if $m_{LL} > m_{RR}$ then
$\cos^2\theta_{\tilde{t}} < \frac{1}{{2}}$ and $\tilde{t}_1$ has a
predominantly right gauge character.

\subsection{Chargino mixing}\label{sec:chamixing}

In the MSSM, the mass matrix of the spin-1/2 partners of the charged
gauge and charged Higgs bosons, $\tilde{W}^+$ and $\tilde{H}^+$,
takes the form
\begin{eqnarray}
{\cal M}_C=\left(\begin{array}{cc}
  M_2       &   \sqrt{2}  m_W\sin\beta \\[2mm]
   \sqrt{2} m_W \cos\beta &     \mu
                  \end{array}\right)\: ,
\label{eq:massmatrix}
\end{eqnarray}
where $M_2$ is the $SU(2)$ gaugino mass parameter. By
reparametrization of the fields, $M_2$ can be taken real and
positive, while the higgsino mass parameter $\mu$ can be complex,
see Eq.~\eqref{eq:at-mu}. Since the chargino mass matrix ${\cal
M}_C$ is not symmetric, two different unitary matrices are needed to
diagonalize it
\begin{eqnarray}
    U^* {\cal M}_C V^\dag =
    \left( \begin{array}{cc}
    m_{\tilde{\chi}^\pm_1} & 0\\
    0 & m_{\tilde{\chi}^\pm_2} \\
    \end{array}\right)\qquad \mbox{with} \qquad m_{\tilde{\chi}^\pm_1} < m_{\tilde{\chi}^\pm_2} \:.
\end{eqnarray}
$U$ and $V$ matrices act on the left- and right-chiral
$\psi_{L,R}=(\tilde{W},\tilde{H})_{L,R}$ two-component states
\begin{eqnarray}
\tilde{\chi}^R_j=U_{jk} \psi_k^R\,, \qquad \tilde{\chi}^L_j=V_{jk}
\psi_k^L\: ,
\end{eqnarray}
giving the two mass eigenstates $\tilde{\chi}^\pm_1$,
$\tilde{\chi}^\pm_2$.

\subsection{Neutralino mixing}

In the MSSM, the four neutralinos $\tilde{\chi}^0_i$ ($i=1,2,3,4$)
are mixtures of the neutral $U(1)$ and $SU(2)$ gauginos, $\tilde{B}$
and $\tilde{W}^3$, and the  higgsinos, $\tilde{H}^0_1$ and
$\tilde{H}^0_2$. The neutralino mass matrix in the $(\tilde{B},
\tilde{W}^3, \tilde{H}^0_1, \tilde{H}^0_2)$ basis,
\begin{eqnarray}
{\cal M}_N =\left(\begin{array}{cccc}
          M_1  &  0  & -m_Z c_\beta s_W   &  m_Z s_\beta s_W  \\[1mm]
          0    & M_2 & m_Z c_\beta c_W    & -m_Z s_\beta c_W  \\[1mm]
       -m_Z c_\beta s_W &  m_Z c_\beta c_W &   0  & -\mu \\[1mm]
        m_Z s_\beta s_W & -m_Z s_\beta c_W & -\mu &  0
               \end{array}\right)
\label{eq:mass_matrix}
\end{eqnarray}
is built up by the fundamental SUSY parameters: the $U(1)$ and
$SU(2)$ gaugino masses $M_1$ and $M_2$, the higgsino mass parameter
$\mu$, and $\tan\beta=v_2/v_1$ ($c_\beta = \cos\beta$, $s_W =
\sin\theta_W$ etc.). In addition to the $\mu$ parameter, a
non-trivial CP phase can also be attributed to the $M_1$ parameter:
\begin{eqnarray}
M_1 = |M_1|\, {\rm e}^{{\rm i}\phi_1},\quad \ \ (0\leq \phi_1 <
2\pi)\,.
\end{eqnarray}
Since the complex matrix ${\cal M}_N$ is symmetric, one unitary
matrix $N$ is sufficient to rotate the gauge eigenstate basis
$(\tilde{B}, \tilde{W}^3, \tilde{H}^0_1, \tilde{H}^0_2)$ to the mass
eigenstate basis of the Majorana fields~$\tilde{\chi}^0_i$
\begin{eqnarray}
{\rm
diag}(m_{\tilde{\chi}_1^0},m_{\tilde{\chi}_2^0},m_{\tilde{\chi}_3^0},m_{\tilde{\chi}_4^0})
= N^*{\cal M}_N N^\dagger\,,\qquad
(m_{\tilde{\chi}_1^0}<m_{\tilde{\chi}_2^0}<m_{\tilde{\chi}_3^0}<m_{\tilde{\chi}_4^0})\,.
\end{eqnarray}
The masses  $m_{\tilde{\chi}_i^0}$ ($i=1,2,3,4$) can be chosen to be
real and positive by a suitable definition of the unitary matrix
$N$.

\subsection{Couplings of stops to charginos and neutralinos}\label{sec:couplings}
We now give explicit formulae for the couplings relevant for decays
Eqs.~\eqref{eq:chab} and \eqref{eq:neu1t} in the convention of
Ref.~\cite{Rosiek:1989rs}. In terms of two-component (Weyl) gauge
eigenstates the coupling between stop, top and neutral
gauginos/higgsinos is given by
\begin{eqnarray}
\mathcal{L} = \mathrm{i} \sqrt{2}\: \tilde{t}^*_L \left(g_2 T^3
\tilde{W}^3 + \frac{1}{6} g_1 \tilde{B} \right) t_L - \mathrm{i}
\frac{2\sqrt{2}}{3} g_1 \tilde{t}^*_R \tilde{B} t_R - Y_t
\tilde{t}^*_R \tilde{H}^0_u t_L - Y_t \tilde{t}^*_L \tilde{H}^0_u
t_R + \mathrm{h.c.}\,,
\end{eqnarray}
where $e=g_2 s_W = g_1 c_W$, $T^3 = \frac{1}{2} \tau^3$ is the
$SU(2)$ generator and $\tau^3$ is the Pauli matrix. After
electroweak symmetry breaking for the mass eigenstates $t$,
$\tilde{t}_i$ and $\tilde{\chi}^0_j$ we get
\begin{eqnarray}\label{eq:neulag}
\mathcal{L}_{t \tilde{t} \tilde{\chi}^0} =
\overline{\tilde{\chi}^0_j} \left( Q^{0,L}_{ij} P_L + Q^{0,R}_{ij}
P_R \right) t \: \tilde{t}^*_i + \mathrm{h.c.} \,,
\end{eqnarray}
where
\begin{eqnarray}
Q^{0,L}_{ij} &=& -\frac{e}{\sqrt{2}\: s_W c_W}\: \mathcal{R}^{\tilde{t}}_{i1} \left( \frac{1}{3} s_W N^*_{j1} + c_W N^*_{j2} \right)  - Y_t\: \mathcal{R}^{\tilde{t}}_{i2} N^*_{j4}\,, \label{eq:q0L}\\
Q^{0,R}_{ij} &=& \frac{2 \sqrt{2}\: e}{3 c_W}\:
\mathcal{R}^{\tilde{t}}_{i2} N_{j1} - Y_t\:
\mathcal{R}^{\tilde{t}}_{i1} N_{j4}\,, \label{eq:q0R}
\end{eqnarray}
with the top Yukawa coupling given by
\begin{eqnarray}
Y_t = \frac{e\: m_t}{\sqrt{2}\: m_W s_W \sin\beta}\,.
\end{eqnarray}
We now see that the right squark couples only to the bino and the
higgsino components of the neutralino. If the $\mu$ parameter is
much larger than the gaugino mass parameters then the light chargino
$\cha^\pm_1$ and light neutralinos $\neu{1},\neu{2}$ are gauginos
with small higgsino components. In this case the Yukawa term in
Eq.~\eqref{eq:q0R} is negligible for stop decays into these states.
On the other hand, as can be seen in Eqs.~\eqref{eq:q0L}
and~\eqref{eq:q0R}, left squarks couple both to the bino and the
wino, however, with the bino coupling suppressed by a factor $1/3$
due to hypercharge. Therefore, having a prior knowledge on the
composition of neutralinos we can infer the structure of the stop
sector by comparing strength of the stop coupling to different
neutralino states.

Let us turn now to the coupling between chargino, stop and bottom
quark. The interaction Lagrangian in terms of gauge eigenstates
reads in Weyl notation
\begin{eqnarray}\label{eq:charginoLPcpl}
\mathcal{L} = \mathrm{i} g_2 \tilde{t}^*_L \tilde{W}^+ b_L + Y_t
\tilde{t}^*_R \tilde{H}_u^+ b_L + Y_b \tilde{t}^*_L \tilde{H}_d^+
b_R + \mathrm{h.c.}
\end{eqnarray}
After electroweak symmetry breaking and rotation of fields to their
mass eigenstates we get
\begin{eqnarray}\label{eq:chalag}
\mathcal{L}_{b \tilde{t} \tilde{\chi}} =
\overline{\tilde{\chi}^{+}_j} \left( Q^{\pm,L}_{ij} P_L +
Q^{\pm,R}_{ij} P_R \right) b\: \tilde{t}^*_i + \mathrm{h.c.}\,,
\end{eqnarray}
where
\begin{eqnarray}
Q^{\pm,L}_{ij} &=& -\frac{e}{s_W}\: \mathcal{R}^{\tilde{t}}_{i1} V^*_{j1} + Y_t\: \mathcal{R}^{\tilde{t}}_{i2} V^*_{j2} \,,\label{eq:qpR}\\
Q^{\pm,R}_{ij} &=& Y_b\: \mathcal{R}^{\tilde{t}}_{i1} U_{j2}
\,,\label{eq:qpL}
\end{eqnarray}
with the bottom Yukawa coupling given by
\begin{eqnarray}
Y_b = \frac{e\: m_b}{\sqrt{2}\: m_W s_W \cos\beta}\,.
\end{eqnarray}
The right stop couples only to the higgsino component of the
chargino via the Yukawa term in Eq.~\eqref{eq:qpL}, whereas the left
stop couples both to the higgsino and the wino. When the light
chargino is mainly wino-like the higgsino couplings are small and
only the stop-bottom-wino term becomes relevant. Therefore,
similarly as for interactions with neutralinos, measurement of
coupling strength to the light chargino can probe the left-right
composition of the light stop.

\section{Stop decays to charginos and neutralinos} \label{sec:decays}
\subsection{Analytical formulae}\label{sec:formulae}

We start this section by giving formulae for the decay widths of the
top squarks into charginos and
neutralinos~\cite{kramlphd,Bartl:2003pd}. Using couplings defined in
Sec.~\ref{sec:couplings} the tree-level width for the decay
Eq.~\eqref{eq:chab} can be written as
\begin{align}\label{eq:chawidth}
\begin{split}
\Gamma(\st_i \to \cha^+_j b) = &
\frac{\kappa(m_{\st_i}^2,m_b^2,m^2_{\cha_j^\pm})}{16 \pi
m_{\st_i}^3}\bigg( \left( \big|Q_{ij}^{\pm,L}\big|^2 +
\big|Q_{ij}^{\pm,R}\big|^2 \right)\left( m_{\st_i}^2 - m_b^2 -
m^2_{\cha_j^\pm} \right)\\ & \hspace*{3.3cm} - 4\, {\rm Re}\left[
Q_{ij}^{\pm,L} Q_{ij}^{\pm,R\, *} \right] m_b m_{\cha_j^\pm}
\bigg)\,,
\end{split}
\end{align}
with the kinematic triangle function
\begin{eqnarray}
\kappa(x,y,z) = \sqrt{(x-y-z)^2 - 4yz}\,,\label{eq:kappa}
\end{eqnarray}
and the couplings $Q^{\pm}_{ij}$ given by Eqs.~\eqref{eq:qpR},
\eqref{eq:qpL}. Substituting the explicit matrix elements of
Eq.~\eqref{eq:stop-mix} we can make the following expansion in terms
of the stop mixing angle and the phase
\begin{align}
\begin{split}\label{eq:chaq2}
\big|Q_{1j}^{\pm,L}\big|^2 + \big|Q_{1j}^{\pm,R}\big|^2 = &
\cos^2\theta_{\st}\: \bigg(Y_b^2 |U_{j2}|^2 + \frac{e^2}{s_W^2}
|V_{j1}|^2\bigg) + \sin^2\theta_{\st}\: Y_t^2 |V_{j2}|^2 \\  &-2
\sin\theta_{\st} \cos\theta_{\st}\:  \frac{e}{s_W} Y_t  \: {\rm Re}
\left[ {\rm e}^{-{\rm i} \phi_{\st}} V_{j1} V_{j2}^*\right]\,,
\end{split}\\
{\rm Re}\left[ Q_{ij}^{\pm,L} Q_{ij}^{\pm,R\, *} \right] = & -
\cos^2\theta_{\st}\: \frac{e}{s_W} Y_b\: {\rm Re} \left[ U_{j2}^*
V_{j1}^* \right]+ \sin\theta_{\st} \cos\theta_{\st}\: Y_b Y_t \:
{\rm Re} \left[  {\rm e}^{-{\rm i} \phi_{\st}} U_{j2}^* V_{j2}^*
\right]\,.\label{eq:chaqrql}
\end{align}
We see explicitly that the dependence of the phase $\phi_{\st}$
appears only if there is a significant higgsino component ($U_{j2}$
or $V_{j2}$) in the chargino $\cha^+_j$ we are interested in.

Analogously, for decays to neutralinos we
have~\cite{kramlphd,Bartl:2003pd}
\begin{align}\label{eq:neuwidth}
\begin{split}
\Gamma(\st_i \to \neu{j} t) = &
\frac{\kappa(m_{\st_i}^2,m_t^2,m^2_{\neu{j}})}{16 \pi
m_{\st_i}^3}\bigg( \left( \big|Q_{ij}^{0,L}\big|^2 +
\big|Q_{ij}^{0,R}\big|^2 \right)\left( m_{\st_i}^2 - m_t^2 -
m^2_{\neu{j}} \right)\\ & \hspace*{3.3cm} - 4\, {\rm Re}\left[
Q_{ij}^{0,L} Q_{ij}^{0,R\, *} \right] m_t m_{\neu{j}} \bigg)\,,
\end{split}
\end{align}
with $\kappa(x,y,z)$ given by Eq.~\eqref{eq:kappa} and couplings
$Q_{ij}^{0}$ by Eqs.~\eqref{eq:q0L}, \eqref{eq:q0R}. Similarly we
obtain
\begin{align}
\begin{split}\label{eq:neuq2}
\big|Q_{1j}^{0,L}\big|^2 + \big|Q_{1j}^{0,R}\big|^2 = &\\ &
\hspace{-3.5cm} = \cos^2\theta_{\st} \left( \frac{e^2}{2 s_W^2 c_W^2
} \Big| \frac{1}{3} s_W N_{j1} + c_W N_{j2} \Big|^2 + Y_t^2
|N_{j4}|^2 \right)
+ \sin^2\theta_{\st} \left(\frac{8 e^2}{9 c_W^2} |N_{j1}|^2 + Y_t^2 |N_{j4}|^2 \right) \\
& \hspace{-3.5cm} + 2 \sin\theta_{\st} \cos\theta_{\st}\: Y_t \left(
\frac{e}{\sqrt{2}\: s_W c_W}\: {\rm Re} \bigg[ {\rm e}^{{\rm i}
\phi_{\st}} \left( \frac{1}{3} s_W N^*_{j1} + c_W N^*_{j2} \right)
N_{j4} \bigg] \right. \left. -\frac{2\sqrt{2}\:e}{3 c_W}\: {\rm Re}
\big[ {\rm e}^{-{\rm i} \phi_{\st}} N_{j1} N_{j4}^* \big] \right)\,
\end{split}\\
\begin{split}\label{eq:neuqlqr}
{\rm Re}\left[ Q_{1j}^{0,L} Q_{1j}^{0,R\, *} \right] = & \\ & \hspace{-2.2cm}=\cos^2\theta_{\st}\: \frac{e}{\sqrt{2}\: s_W c_W} Y_t\: {\rm Re} \left[\left( \frac{1}{3} s_W N^*_{j1} + c_W N^*_{j2} \right) N_{j4}^* \right] + \sin^2\theta_{\st}\: \frac{2\sqrt{2}\: e}{3 c_W}\: Y_t\: {\rm Re}[N_{j4}^* N_{j1}^*]\\
& \hspace{-2.2cm}+ \sin\theta_{\st} \cos\theta_{\st} \left( Y_t^2\:
{\rm Re} \left[{\rm e}^{-{\rm i} \phi_{\st}} N_{j4}^{*2} \right] -
\frac{2}{3} \frac{e^2}{s_W c_W^2}\: {\rm Re} \left[  {\rm e}^{{\rm
i} \phi_{\st}} \left( \frac{1}{3} s_W N^*_{j1} + c_W N^*_{j2}
\right) N_{j1}^* \right] \right)\,.
\end{split}
\end{align}

An interesting feature of Eqs.~\eqref{eq:chawidth} and
\eqref{eq:neuwidth} is the relative importance of the squared
$|Q_{ij}^{L}\big|^2 + \big|Q_{ij}^{R}\big|^2$ terms and the
left-right interference ${\rm Re} \left[Q_{ij}^{L} Q_{ij}^{R\, *}
\right]$ terms. As they are multiplied by mass factors, it is going
to be sensitive to the mass splitting between stop and $\cha^+_i\,
b$, $\neu{i}\, t$ pairs. If the given decay mode is close to its
kinematic threshold (which will be the case for heavier neutralinos)
the second term will become dominant, whereas far from the threshold
the first term will usually be much larger.

In the above discussion, higher-order effects have been neglected.
Different parts of the one-loop corrections have been calculated by
many groups and the full one-loop result for the branching ratios in
the real MSSM can be found in e.g.~\cite{full1loop}. The size of the
corrections depends on the scenario and can exceed 10\%. These
corrections will of course depend on the full MSSM parameter set.
Moreover, depending on the renormalization scheme applied, one will
have to redefine the stop mixing angle accordingly,
cf.~\cite{Hollik:2003jj}. This, however, does not limit the
applicability of the method presented here. Thus a determined mixing
angle can be treated as an effective parameter. In order to extract
a one-loop corrected mixing angle, one will have to know the
dominant SUSY corrections to $\tilde{t}$ decays. This could be
possible with high integrated luminosity at the LHC. The most
accurate prediction of the model parameters will be made when a
global SUSY fit is performed with many different observables. For
the stop sector, the proposed ratio of branching ratios method could
provide useful additional information for these fits. For an
ultimate precision, however, a future linear collider will be needed
(see Ref.~\cite{Weiglein:2004hn} and references therein).

\subsection{Discussion of typical mixing scenarios}

In order to analyze the dependence of the stop mixing angle on the
decay widths and the branching ratios, we consider three benchmark
points of the MSSM. The first scenario is the well known mSUGRA
inspired SPS1a$'$ parameter point~\cite{AguilarSaavedra:2005pw} --
in the following we will refer to it as Scenario A. A feature of
mSUGRA scenarios is that the charginos and the neutralinos are to a
large extent pure gaugino/higgsino states: the lightest neutralino
is bino-like, the light chargino and the second neutralino are
winos, and the heavy chargino and the heavy neutralinos are
higgsino-like. Scenarios B and C are adopted from
Ref.~\cite{Ellis:2008hq}. In Scenario B the wino mass parameter
$M_2$ and the higgsino mass parameter $\mu$ are of a similar order,
giving strong mixing between the wino and the higgsino components of
the charginos and the neutralinos. This makes the determination of
$\theta_{\st}$ more difficult since both left and right couplings of
Eqs.~\eqref{eq:neulag} and~\eqref{eq:chalag} contribute to all the
final states considered. On the other hand this gives the
possibility to study the dependence on the CP-violating phase
$\phi_{\st}$, thanks to the last terms of Eqs.~\eqref{eq:chaq2},
\eqref{eq:chaqrql}, \eqref{eq:neuq2} and~\eqref{eq:neuqlqr}.
Finally, Scenario C features the wino mass parameter two times
larger than the $\mu$ parameter. In this case higgsino-like states
will be lighter than winos with rather small mixing. In both cases,
Scenarios B and C, the lightest supersymmetric particle
$\tilde{\chi}_1^0$ is bino-like. In order to study the possible
dependence of branching ratios on the CP-violating phase in the last
two scenarios we introduce a CP phase for the stop trilinear
coupling $A_t$. For all three scenarios we keep the values of other
parameters (i.e.\ slepton and squark sectors) as in the SPS1a$'$
scenario. The values of the gaugino, higgsino and stop sector
parameters are collected in Tab.~\ref{tab:mssm} and the nominal
values of masses, mixing angles and branching ratios are listed in
Tables~\ref{tab:masses} and~\ref{tab:brs}.

\begin{table} \renewcommand{\arraystretch}{1.3}
\begin{center}
\begin{tabular}{|c||c|c|c|c|c|c|c|}\hline
           & $M_1$ & $M_2$ & $\mu$  & $\tan\beta$ & $A_t$                  & $M_{Q_3}$ & $M_{U_3}$ \\ \hline\hline
Scenario A & 103.3 & 193.2 & 396.0  &          10 & $-565.1$
& 471.4 & 387.5 \\ \hline Scenario B & 109.0 & 240.0 & 230.0  &
10 & $-610\:\mathrm{e}^{\mathrm{i}\: \pi/2}$& 511.0 & 460.0 \\
\hline Scenario C & 105.0 & 400.0 & $-190.0$ &        20 &
$-610\:\mathrm{e}^{\mathrm{i}\: \pi/4}$& 511.0 & 460.0 \\ \hline
\end{tabular}
\caption{MSSM parameters of Scenarios A, B and C relevant in the
present study. Mass parameters and trilinear coupling are given in
GeV. \label{tab:mssm}}
\end{center}
\end{table}

\begin{table}[t!] \renewcommand{\arraystretch}{1.3}
\begin{center}
\begin{tabular}{|c||c|c||c|c||c|c|c|c|} \hline
          & $m_{\st_1}$ & $m_{\st_2}$ & $\mcha{1}$ & $\mcha{2}$ & $\mneu{1}$ & $\mneu{2}$ & $\mneu{3}$ & $\mneu{4}$ \\[0.16em]\hline\hline
Scenario A&     366.5   &   585.5 & 183.7      & 415.7  & 97.7
& 183.9      & 400.5      & 413.9      \\\hline Scenario B&
395.5   &   609.0     & 178.0      & 302.9      & 101.4      & 182.2
& 237.8      & 303.2      \\\hline Scenario C&     396.9   &   608.0
& 182.9      & 419.0      & 99.0       & 186.6      & 199.5      &
418.9      \\\hline
\end{tabular}
\caption{Masses (in GeV) of stops, charginos and neutralinos in
Scenarios A, B and C calculated by \texttt{SPheno
2.2.3}~\cite{spheno}. \label{tab:masses}}
\end{center}
\end{table}

\begin{table}[t!] \renewcommand{\arraystretch}{1.3}
\begin{center}
\begin{tabular}{|c||c|c|c|} \hline
 Parameter                     & Scenario A & Scenario B & Scenario C \\ \hline \hline
 $\cos \theta_{\st}$           & 0.56       &   0.62     & 0.62       \\ \hline
 $\phi_{\st}$                  & 0          &   1.53     & 0.80       \\ \hline
 $\Gamma(\st_1)$ [GeV]         & 1.45       & 3.25       & 6.36       \\ \hline
 $BR(\st_1 \to \cha_1^+ b)$    & 73.5\%     & 60.7\%     & 63.7 \%    \\ \hline
 $BR(\st_1 \to \cha_2^+ b)$    & ---        & 17.6\%     & ---        \\ \hline
 $BR(\st_1 \to \neu{1} t)$     &  20.1\%    &  8.8\%     & 6.4\%      \\ \hline
 $BR(\st_1 \to \neu{2} t)$     &  6.4\%     & 12.9\%     & 8.5\%      \\ \hline
 $BR(\st_1 \to \neu{3} t)$     & ---        & ---        & 21.4\%     \\ \hline
 $\sigma(pp\to \st_1 \st_1^*)$ [pb] &  3.44 & 2.27       & 2.27       \\ \hline
\end{tabular}
\caption{Nominal values of mixing angles in the stop sector and
branching ratios for stop decays calculated from
Eqs.~\eqref{eq:chawidth} and \eqref{eq:neuwidth} for Scenarios A, B
and C. In the last row, cross sections for stop pair production at
the LHC with $\sqrt{s} = 14$~TeV at NLO from \texttt{Prospino
2.1}~\cite{beenakker98,prospino}.\label{tab:brs}}
\end{center}
\end{table}

We now discuss the behaviour of the decay widths and the branching
ratios with respect to the stop mixing angle and the CP phase in
each of the scenarios.

\subsubsection*{Scenario A -- mSUGRA}
\begin{figure}[t!]
\begin{center}
\includegraphics[width=0.47\textwidth]{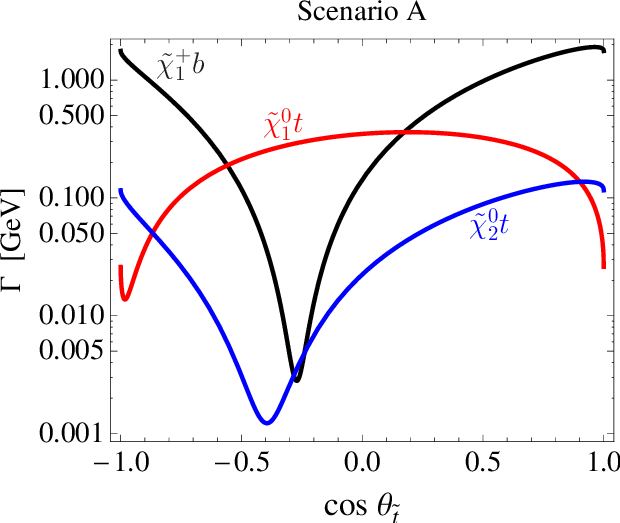} \hskip 0.5cm
\includegraphics[width=0.45\textwidth]{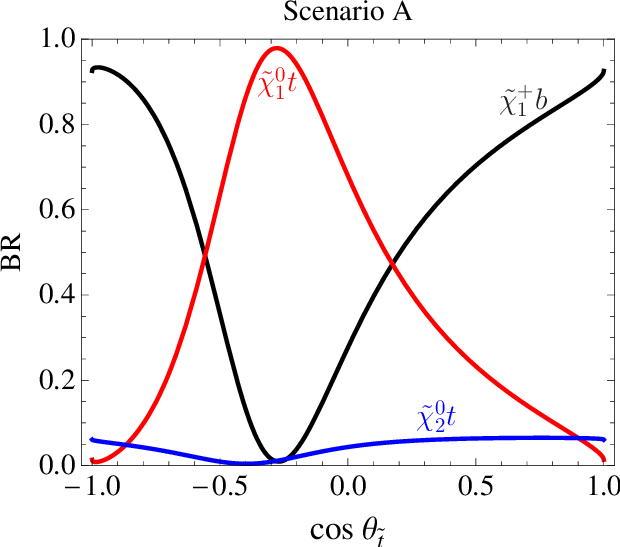} \vskip 0.5cm
\includegraphics[width=0.45\textwidth]{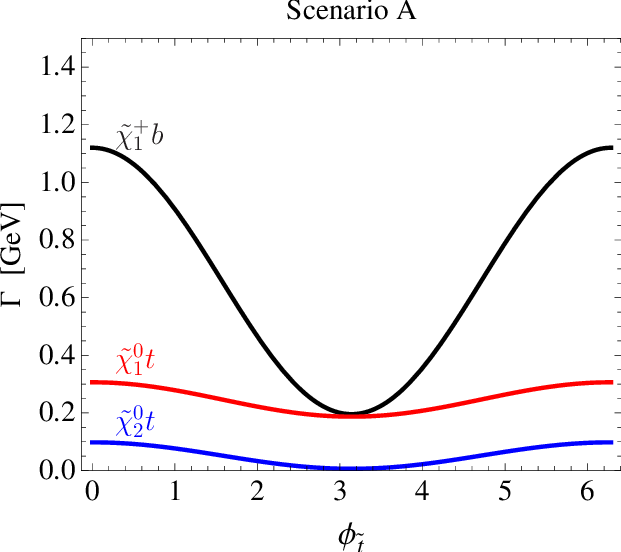} \hskip 0.5cm
\includegraphics[width=0.45\textwidth]{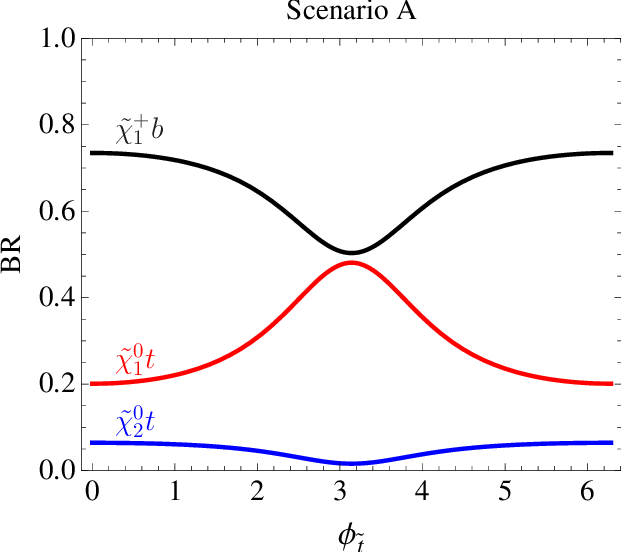}
\caption{Decay widths (left column) and branching ratios (right
column) for $\st_1$ in Scenario A as a function of the stop mixing
angle $\cos\theta_{\st}$ (upper row) and the stop CP phase
$\phi_{\st}$ (lower row). Black, red and blue lines are for the
$\tilde{\chi}^+_1 b$, $\neu{1}\, t$ and $\neu{2}\, t$ final states,
respectively. \label{fig:scA-gamma}  }
\end{center}
\end{figure}

According to the discussion in Sec.~\ref{sec:couplings}, for
Scenario A we expect that if the $\tilde{t}_1$ is mainly a left stop
(i.e.\ for $\cos\theta_{\st} = \pm 1$) then it will dominantly
couple to $\tilde{\chi}^+_1$ and $\neu{2}$ (which are both winos),
whereas the coupling to the bino-like $\neu{1}$ is suppressed. On
the other hand, if the $\tilde{t}_1$ is predominantly a right stop
(i.e.\ for  $\cos\theta_{\st} = 0$) we should observe enhancement in
the coupling to the LSP and suppression for the decay to the light
chargino and second neutralino. This general feature can be seen in
the upper left panel of Fig.~\ref{fig:scA-gamma}, where we show the
dependence of the decay width on the stop mixing angle
$\cos\theta_{\st}$. The minima for decays to chargino and $\neu{2}$
are somewhat shifted which is the result of the higgsino Yukawa
contributions from Eqs.~\eqref{eq:q0L}, \eqref{eq:q0R},
\eqref{eq:qpR} and \eqref{eq:qpL}. On top of that, the decay $\st_1
\to \neu{2}\, t$ is further suppressed by the phase space, since
$\mneu{2} + m_t = 355$~GeV is only slightly lower than the light
stop mass. As one can see, the decay widths change by an order of
magnitude or more. Therefore they are a sensitive probe of the
mixing between left and right stop states. The upper right panel of
Fig.~\ref{fig:scA-gamma} shows the dependence of the branching
ratios on $\cos\theta_{\st}$ that exhibit a similar behaviour as the
decay widths.

Although Scenario A does not contain CP phases, we include them here
to analyze the sensitivity of the decay widths and the branching
ratios. The respective plots can be seen in the lower row of
Fig.~\ref{fig:scA-gamma}. The most significant change is for the
decay to a chargino and a bottom quark. This results from the third
term of Eq.~\eqref{eq:chaq2} that changes sign when varying
$\phi_{\st}$ from 0 to $\pi$ giving destructive interference.
Although the dependence on $\phi_{\st}$ is clearly visible the
constraints on this parameter, as we will see it later, will be
rather weak.

\subsubsection*{Scenario B -- mixed gaugino/higgsino $\cha_1^+$ and $\neu{2}$}
\begin{figure}[t!]
\begin{center}
\includegraphics[width=0.45\textwidth]{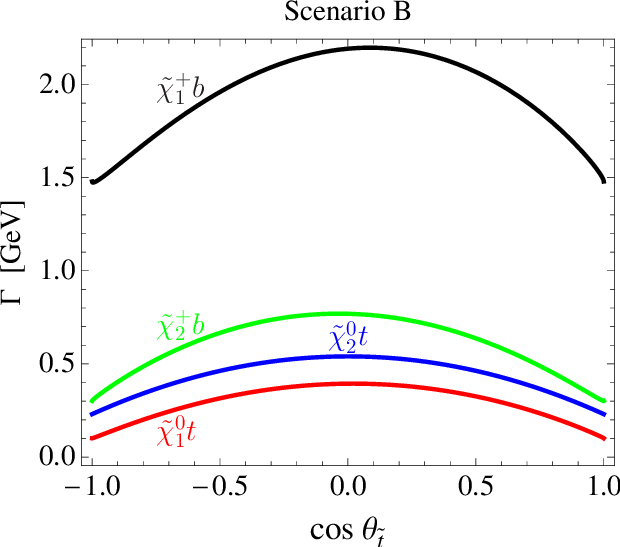} \hskip 0.5cm
\includegraphics[width=0.45\textwidth]{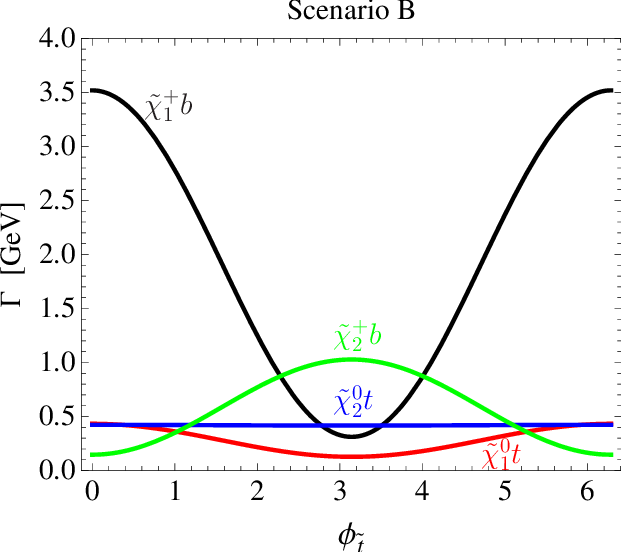}
\caption{Decay widths for $\st_1$ in Scenario B as a function of the
stop mixing angle $\cos\theta_{\st}$ (left panel) and the stop CP
phase $\phi_{\st}$ (right panel). Black, red, blue and green lines
are for the $\tilde{\chi}^+_1 b$, $\neu{1}\, t$, $\neu{2}\, t$ and
$\tilde{\chi}^+_2 b$ final states, respectively.
\label{fig:scB-gamma}  }
\end{center}
\end{figure}

The situation changes significantly in Scenario B. The second
chargino $\cha^+_2$ is now lighter than the stop $\st_1$ so there is
a new decay channel open. Both charginos and the neutralino
$\neu{2}$ now have a significant higgsino component. The dependence
on the stop mixing angle of all the decay widths is much flatter
now, as can be seen in the left panel of Fig.~\ref{fig:scB-gamma}.
We note a well pronounced enhancement of the decay widths for right
stops (around $\cos\theta_{\st} = 0$). For charginos it can be
understood by looking at Eq.~\eqref{eq:charginoLPcpl} where the
coupling of $\st_R$ is proportional to the large top Yukawa
coupling, whereas the coupling of $\st_L$ is proportional to the
smaller bottom Yukawa coupling. For the decay to $\neu{2}\,t$ the
enhancement is due to the left-right interference term, whereas for
$\neu{1}\,t$ it is due to the quadratic term in
Eq.~\eqref{eq:neuwidth}.

Thanks to the presence of the higgsino component in charginos and
neutralinos, we now become more sensitive to the phase $\phi_{\st}$,
see the right panel of Fig.~\ref{fig:scB-gamma}. Apart from the
decay to $\neu{2}\,t$, all other decay widths can change by up to an
order of magnitude depending on the CP phase. The former remains
almost unchanged due to the accidental cancellations between two
terms of Eq.~\eqref{eq:neuwidth}. For the decays to charginos
$\cha_1^\pm$ ($\cha_2^\pm$) the suppression (enhancement) of the
decay width with the phase arises due to change of the sign of
$\cos\phi_{\st}$ when $\phi_{\st} \to \pi$, cf.\
Eqs.~\eqref{eq:chaq2} and~\eqref{eq:chaqrql}. The difference for
$\cha_1^+$ and $\cha_2^+$ is the result of the sign difference
between $V_{11} V_{12}^*$ and $V_{21} V_{22}^*$.

\begin{figure}[t!]
\begin{center}
\includegraphics[width=0.45\textwidth]{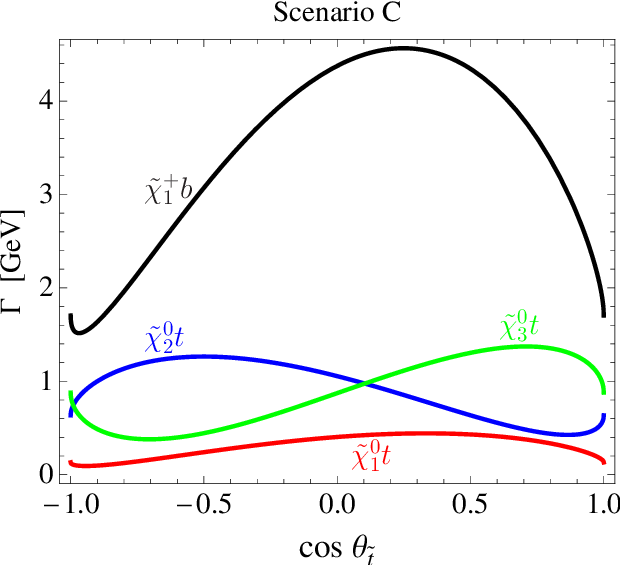} \hskip 0.5cm
\includegraphics[width=0.45\textwidth]{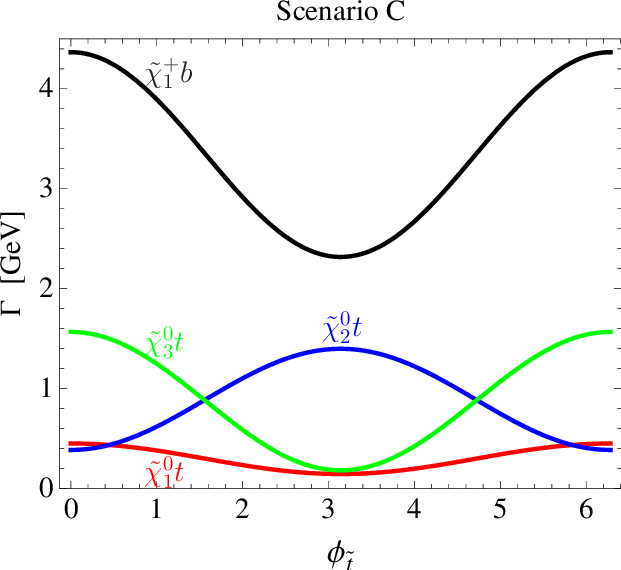}
\caption{Decay widths for $\st_1$ in Scenario C as a function of the
stop mixing angle $\cos\theta_{\st}$ (left panel) and the stop CP
phase $\phi_{\st}$ (right panel). Black, red, blue and green lines
are for the $\tilde{\chi}^+_1 b$, $\neu{1}\, t$, $\neu{2}\, t$ and
$\neu{3}\,t$ final states, respectively. \label{fig:scC-gamma}  }
\end{center}
\end{figure}

\subsubsection*{Scenario C -- higgsino-like $\cha_1^+$, $\neu{2}$ and $\neu{3}$}
The last discussed scenario features the hierarchy $M_1 < \mu <
M_2$.  Therefore the light chargino and neutralinos $\neu{2}$,
$\neu{3}$ are higgsino-like with small mass differences between
them. The lightest neutralino is bino-like as in the previous
scenarios. The dependence of the decay widths on the stop mixing
angle has been shown in the left panel of Fig.~\ref{fig:scC-gamma}.
The difference in the decay widths to the chargino for left and
right stops is a consequence of the $Y_t$ coupling for right states
in Eq.~\eqref{eq:charginoLPcpl}. A similar effect was seen in
Scenario B, however it is now more pronounced due to the higgsino
nature of the light chargino $\cha_1^+$. We also observe the
interesting exchange of the decay widths to heavier neutralinos when
the sign of $\cos\theta_{\st}$ changes. This feature arises due to
the $ Y_t^2\: {\rm Re} \left[ N_{j4}^{*2} \right]$ term in the
second line of Eq.~\eqref{eq:neuqlqr} that is enhanced both by the
large top Yukawa coupling and the higgsino nature of the two
neutralinos. Since neutralinos $\neu{2}$ and $\neu{3}$ have opposite
intrinsic CP parities in Scenario~C, cf.\ Ref.~\cite{Choi:2001ww},
the entries in the neutralino mixing matrix that correspond to
$\neu{3}$ are purely imaginary. Therefore, the contribution has an
opposite sign in the decay width and hence different behaviour with
respect to the sign of $\cos\theta_{\st}$.

A similar dependence of the decay widths $\neu{2}\,t$ and
$\neu{3}\,t$ on the sign of $\cos\phi_{\st}$ can be seen in the
right panel of Fig.~\ref{fig:scC-gamma}. Its origin is the same as
in the above discussed case for $\cos\theta_{\st}$. As before the
change in the width of the decay to $\cha^+_1 b$ is caused by a
change in the sign of the last term of Eq.~\eqref{eq:chaq2} with
$\cos\phi_{\st}$, as $\phi_{\st}$ is varied from 0 to $\pi$. It is
interesting to note that now the width for the decay to chargino
$\cha_{1}^+$ does not show as strong dependence on the phase
$\phi_{\st}$ compared to Scenario~B. However, the dependence of the
branching ratios for the decays to neutralinos is still well
pronounced.

\section{Potential observables at the LHC}\label{sec:results}
\subsection{Ratios of branching ratios}
\begin{figure}
\begin{center}
\includegraphics[width=0.45\textwidth]{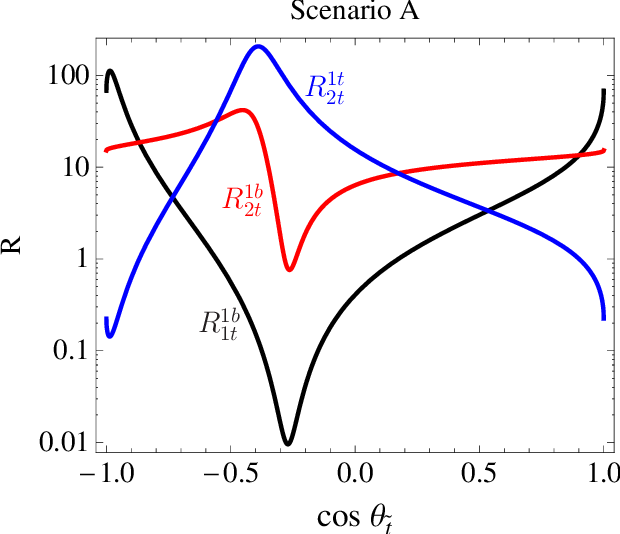} \hskip 0.5cm
\includegraphics[width=0.45\textwidth]{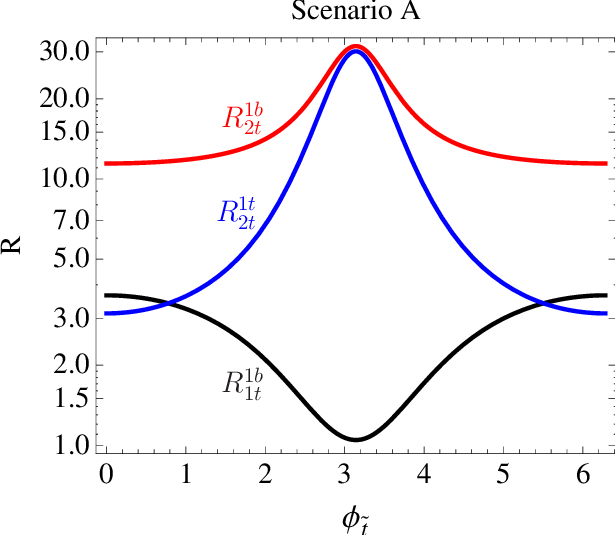}
\caption{Ratios of branching ratios as a function of stop mixing
angle $\cos\theta_{\st}$ and CP phase $\phi_{\st}$ in Scenario A,
left and right panel, respectively. Black, red and blue lines are
for the ratios ${R}_{1t}^{1b}$, ${R}_{2t}^{1b}$ and ${R}_{2t}^{1t}$
(see Eq.~\eqref{eq:ratios}), respectively.\label{fig:ratios-scA}}
\end{center}
\end{figure}

As one can see in Figs.~\ref{fig:scA-gamma}, \ref{fig:scB-gamma} and
\ref{fig:scC-gamma}, the decay widths can change by up to a few
orders of magnitude depending on the stop mixing angle and the CP
phase. In addition, the branching ratios are also very sensitive to
these parameters. However, since the measurement of decay widths and
branching ratios will be difficult at the LHC we propose to analyze
the ratios of branching ratios. That means comparing the number of
stops decaying to one final state with the number of stops decaying
to another final state. Having three decay modes possible we can
define the following ratios of branching ratios for each of the
Scenarios A, B and C:
\begin{eqnarray} \label{eq:ratios}
R_{1t}^{1b} = \frac{BR(\st_1 \to \cha^+_1 b)}{BR(\st_1 \to
\neu{1}\,t)},\qquad R_{2t}^{1b} = \frac{BR(\st_1 \to \cha^+_1
b)}{BR(\st_1 \to \neu{2}\,t)}, \qquad R_{2t}^{1t} = \frac{BR(\st_1
\to \neu{1}\, t)}{BR(\st_1 \to \neu{2}\, t)}\,.
\end{eqnarray}
Figure~\ref{fig:ratios-scA} shows the above ratios of branching
ratios in Scenario A as functions of $\cos\theta_{\st}$ and the
CP-violating phase $\phi_{\st}$. For Scenario B we have three
additional combinations due to the decay $\st_1 \to \cha^+_2 b$
being open,
\begin{eqnarray} \label{eq:ratiosB}
R_{2b}^{1b} = \frac{BR(\st_1 \to \cha^+_1 b)}{BR(\st_1 \to \cha^+_2
b)},\qquad R_{2b}^{1t} = \frac{BR(\st_1 \to \neu{1}\, t)}{BR(\st_1
\to \cha^+_2 b)}, \qquad R_{2b}^{2t} = \frac{BR(\st_1 \to \neu{2}\,
t)}{BR(\st_1 \to \cha^+_2 b)}\, .
\end{eqnarray}
For Scenario C due to the decay $\st_1 \to \neu{3}\,t$ being allowed
we have
\begin{eqnarray} \label{eq:ratiosC}
R_{3t}^{1b} = \frac{BR(\st_1 \to \cha^+_1 b)}{BR(\st_1 \to
\neu{3}\,t)},\qquad R_{3t}^{1t} = \frac{BR(\st_1 \to \neu{1}
t)}{BR(\st_1 \to \neu{3}\,t)}, \qquad R_{3t}^{2t} = \frac{BR(\st_1
\to \neu{2}\, t)}{BR(\st_1 \to \neu{3}\, t)}\,.
\end{eqnarray}
Because of the higgsino nature of neutralinos $\neu{2}$ and
$\neu{3}$ they are very close in mass and it might turn out that
they are impossible to disentangle at the LHC. Therefore we define
two additional ratios by combining the decay modes to $\neu{2}\, t$
and $\neu{3}\, t$
\begin{eqnarray}\label{eq:ratiosC1}
R_{23t}^{1b} = \frac{BR(\st_1 \to \cha^+_1 b)}{BR(\st_1 \to
\neu{2}\,t) + BR(\st_1 \to \neu{3}\,t)}\:,\qquad R_{23t}^{1t} =
\frac{BR(\st_1 \to \neu{1} t)}{BR(\st_1 \to \neu{2}\,t) + BR(\st_1
\to \neu{3}\,t)}\,.
\end{eqnarray}

In our analysis we focus on direct stop production  $pp \to \st_1
\st_1^*$ in order to have better control on the number of observed
stops and to reduce the background due to bottom squarks. In the
SPS1a$'$ scenario the cross section for this process amounts to
3.44~pb at the next-to-leading order~\cite{beenakker98,prospino},
whereas the total SUSY cross section is 60~pb.\footnote{The cross
sections for stop pair production in Scenarios B and C are given in
Tab.~\ref{tab:brs}.} Due to mass splitting between stop states the
cross section for $pp \to \st_2 \st^*_2$ is much smaller with a
value of 0.26~pb. Similarly for sbottoms we get $\sigma(pp \to
\sbot_1 \sbot_1^*) = 0.6$~pb and $\sigma(pp \to \sbot_2 \sbot_2^*) =
0.4$~pb. This gives a relatively clean environment for the
observation of direct light stop pair production and further
reduction after cuts may be expected. Possible final states are as
follows:
\begin{eqnarray}
&& pp \to \st_1 \st_1^* \to t \neu{2} + \bar{t} \neu{2} \to t \ell^+ {\ell^-} \neu{1} + \bar{t} \ell^+ {\ell}^- \neu{1} \to 4\ell\: 4j\: 2b\ + E_{\rm miss} \, , \label{eq:neu2neu2}\\
&& pp \to \st_1 \st_1^* \to t \neu{2} + \bar{t} \neu{1} \to t \ell^+ {\ell}^- \neu{1} + \bar{t} \neu{1} \to 2\ell\: 4j\: 2b + E_{\rm miss}  \, , \\
&& pp \to \st_1 \st_1^* \to t \neu{2} + b \cha^+_1 \to t \ell^+ {\ell}^- \neu{1} + b {\ell}^+ \nu_\ell \neu{1} \to 3\ell\: 2j\: 2b + E_{\rm miss}\, , \\
&& pp \to \st_1 \st_1^* \to t \neu{1} + \bar{t} \neu{1} \to 4j\: 2b + E_{\rm miss}\, , \label{eq:tlsp}\\
&& pp \to \st_1 \st_1^* \to t \neu{1} + b \cha^+_1 \to t \neu{1} + b {\ell}^+ \nu_\ell \neu{1} \to \ell\: 2j\: 2b \: + E_{\rm miss}\, ,\\
&& pp \to \st_1 \st_1^* \to \bar{b} \cha^-_1 + b \cha^+_1 \to
\bar{b} \ell^- \bar{\nu}_\ell \neu{1} + b {\ell}^+ \nu_\ell \neu{1}
\to 2\ell\: 2b \: + E_{\rm miss}\, .\label{eq:bchabcha}
\end{eqnarray}

The production process itself can be tagged using a clean decay mode
for one of the stops, for instance the decay to $\neu{2}\,t$
followed by a leptonic neutralino decay and hadronic top decay. For
an integrated luminosity of $\mathcal{L} = 100\; \mathrm{fb}^{-1}$
we would have more than $300\,000$ stop pair production events in
the case of SPS1a$'$ scenario. Assuming that on average $10\%$ of
charginos and neutralinos decay to leptons~\cite{spheno}, taking
into account the hadronic top branching ratio and varying a
selection efficiency of $1\%$, $3\%$ and $5\%$, one can expect
roughly 350, 1000 and 1750 events to be observed, respectively. A
further increase in the integrated luminosity will result in larger
samples. Therefore in our further analysis we will study the case
when 1000 events have been correctly identified and show that with
this amount of experimental data one can still get strong
constraints on the stop mixing angle and the mass.

The other important point we wish to emphasize are the branching
ratios for decays of the chargino $\cha^\pm_1$ and the neutralino
$\neu{2}$ into leptons. Although one may expect that the related
uncertainty will cancel out to some extent in the ratio
$R^{1b}_{2t}$ (as in our scenarios $\cha^\pm_1$ and $\neu{2}$ have
similar gaugino/higgsino composition), this is not true for the
other ratios involving decays to the LSP. Since our focus here is on
the stop sector we assume that the leptonic branching ratios of the
charginos and neutralinos are known. However, as this would require
a better knowledge of the gaugino/higgsino structure, it is possible
that the measurements from the LHC would have to be supplemented by
the linear collider experiment. Here charginos and neutralinos can
be measured with a high precision $\mathcal{O}(1\%)$, see
Ref.~\cite{Desch:2003vw} and references therein. This would be an
interesting example of LHC/ILC interplay~\cite{Weiglein:2004hn}, in
particular for the scenarios where direct stop production is beyond
the kinematical reach of the ILC.

A large number of SUSY and SM backgrounds are expected for stop
production at the LHC. The most severe Standard Model background,
especially for the channels
Eqs.~\eqref{eq:tlsp}--\eqref{eq:bchabcha}, will be $t\bar{t}$
production. The issue of whether these signals can be distinguished
from the background is under discussion and requires further
experimental studies. For instance, the analysis in
Ref.~\cite{Meade:2006dw} shows that within certain region of stop
masses discovery looks promising in the $t\bar{t}+E_{\mathrm{miss}}$
channel with fully hadronic top decays. A more recent
study~\cite{Plehn:2010st}, using the top-tagging~\cite{Plehn:2009rk}
technique showed that the SM $t\bar{t}$ background can be manageable
for the final state Eq.~\eqref{eq:tlsp}.

The most important SUSY background process is going to be gluino
production with subsequent decays to stops or sbottoms. One
important difference between the signal and these backgrounds is the
number of $b$-jets. The signal event always results in exactly 2
$b$-jets, whereas SUSY backgrounds will typically have 4 $b$-jets
and this feature can be used to suppress them. On the other hand,
squark production followed by decay to chargino and leptonic
chargino decay will also result in 2 jets and 2 leptons in the final
state with a rather high rate. This can fake the signal process
Eq.~\eqref{eq:bchabcha}. However, the background will have two light
quark jets instead of $b$-jets. Therefore, in order to extract the
signal, good $b$-jet tagging efficiency and light jet rejection will
be needed~\cite{Aad:2009wy}. Requiring at least one and no more than
two $b$-jets to be tagged would suppress the backgrounds
significantly.

Another method of getting a handle on the supersymmetric and the
Standard Model backgrounds has been discussed
in~\cite{MoortgatPick:2010wp}. Using kinematic reconstruction, it
was shown that for the final states containing $\neu{2}\to
\ell^+\ell^-\neu{1}$ one can suppress the SUSY backgrounds to the
signal level and, at the same time, practically remove the SM
contributions. This method gave a $S/B$ ratio of order 1, with an
efficiency of about 5\% depending upon the scenario. This is an
encouraging result and the application to other final states,
Eqs.~\eqref{eq:neu2neu2}--\eqref{eq:bchabcha}, will be investigated.

Finally, we note that the signal process with leptonic top decay,
e.g.\ $\neu{1}\,t \to b \ell + E_{\rm miss}$, can give the same
final state as the decay mode with charginos, i.e.\ $\cha^+_1 b \to
b \ell + E_{\rm miss}$. However, we note that this complication does
not affect the result of the fit since it does not introduce any new
unknown parameters. The fitted observables would be a linear
combination of the original ones, Eq.~\eqref{eq:ratios}, and the fit
would rely on the same set of information. Hence, one can combine
the above channels and actually enhance the signal.

An important note is that it will not be sufficient to simply remove
as much background as possible using the relevant cuts. We will also
need to understand with a high degree of accuracy how each
individual signal channel will be affected by the backgrounds.
Understanding the background well is required, as for each channel
we study, the number of background events contaminating the sample
will be different. Therefore the pollution due to backgrounds will
affect our ratios of branching ratios. The reconstruction
efficiency, cuts and triggers will also have a different effect on
each channel and will have to be well understood for our
measurements to be accurate. Another complication will be
higher-order QCD effects, in particular in case of hadronic top
decays. However, the study of this type will require high integrated
luminosity and after a long running time these issues should be much
better understood. We leave a detailed analysis of these effects for
our different final states and the additional uncertainties for
future work.

To summarize, a successful application of the method presented here
will require a good understanding of the detectors and backgrounds.
This may become possible after a few years of the LHC operating in
the high luminosity mode. Also, new analysis techniques that are
being constantly developed should improve our control of the
backgrounds. In line with the results presented
in~\cite{Plehn:2010st,MoortgatPick:2010wp} we therefore assume that
a signal-to-background ratio close to 1 can be achieved. This will
lead to the increased error on the measured number of events by a
factor $\sqrt{3}$. In the following, we take the slightly more
conservative assumption that the error is twice that of the
statistical error alone. A more reliable estimate will require a
complete simulation of signal and background processes.

\subsection{Determination of stop mass and mixing angle}

In order to show the possible advantages of using ratios of
branching ratios for the analysis of the stop sector we first define
the normalized ratios
\begin{eqnarray}\label{eq:normalized}
\widehat{R}_j^i = R_j^i - R^{i\:\mathrm{nominal}}_j\quad
\mbox{with}\quad R^{i\:\mathrm{nominal}}_j = R_j^i(\cos
\theta^{\mathrm{nominal}}_{\st})\,,
\end{eqnarray}
where $\theta^{\mathrm{nominal}}_{\st}$ is the actual mixing angle
in the given scenario and $i$, $j$ run over all possible channels,
i.e.\ $1b$, $1t$, $2t$ etc., cf.\
Eqs.~\eqref{eq:ratios}--\eqref{eq:ratiosC1}. According to this
definition $\widehat{R}_i(\cos
\theta^{\mathrm{nominal}}_{\st})\equiv 0$. Furthermore, we assume
that we have $n=1000$ of well identified events of stop $\st_1
\st_1^* $ pair production. We now take the expected number of events
in each decay mode $n_i = n \times BR_i$. Note that $1 = \sum n_i$
only if decays to charginos and neutralinos are the only possible
decay channels. One should be aware that for our method it is not
necessary to measure all possible decay modes, as explained later.
The statistical error for $n_i$ is $(\Delta n_i)_{\mathrm{stat}} =
\sqrt{n_i}$. In order to account for some of the expected additional
uncertainties, e.g.\ due to background subtraction, in the following
$\chi^2$ fits we decide to use twice as large error, i.e.
\begin{equation}\label{eq:error}
\Delta n_i =2 (\Delta n_i)_{\mathrm{stat}} = 2\sqrt{n_i}\;.
\end{equation}
The resulting error for ratios of branching ratios is given by
\begin{eqnarray}\label{eq:dri}
\Delta R_j^i =  \sqrt{\left(\frac{\Delta n_i}{n_j} \right)^2 +
\left( \frac{n_i \Delta n_j}{n_j^2} \right)^2}\qquad
\mathrm{with}\qquad R_j^i = \frac{n_i}{n_j}\:.
\end{eqnarray}

\begin{figure}
\begin{center}
\includegraphics[width=0.47\textwidth]{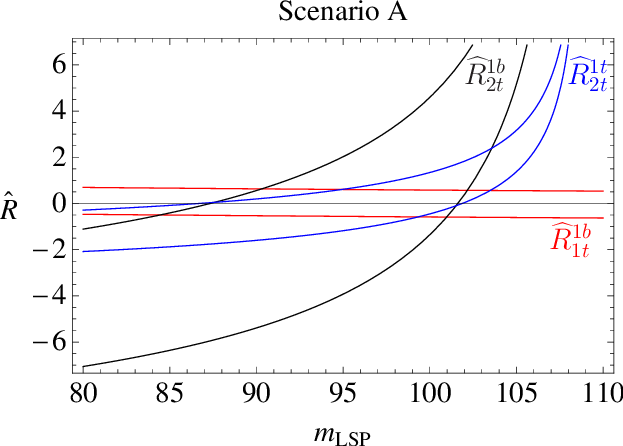}\hskip 0.5cm
\includegraphics[width=0.47\textwidth]{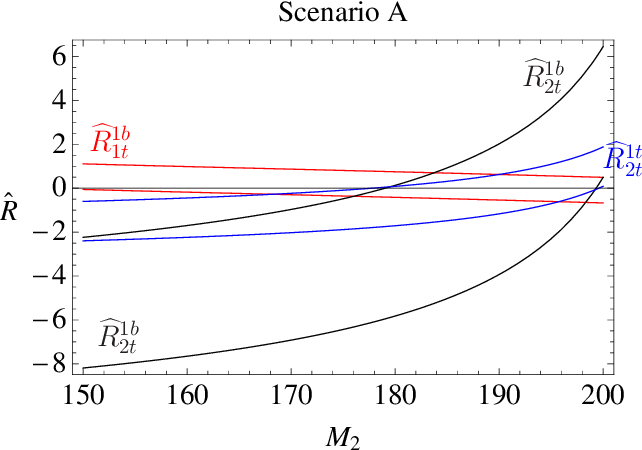}
\caption{Dependence of the normalized ratios of branching ratios
$\widehat{R}_i$, Eq.~\eqref{eq:normalized}, in stop decays on the
mass of the lightest neutralino (left panel) and the gaugino mass
parameter $M_2$ (right panel) in Scenario~A. The values of the other
parameters and the mass difference between the LSP and $\neu{2} $,
$\cha_1^\pm$ for the left plot are fixed to their nominal values.
Red lines are for $\widehat{R}_{1t}^{1b}$, black lines for
$\widehat{R}_{2t}^{1b}$, and blue lines for $\widehat{R}_{2t}^{1t}$,
see Eq.~\eqref{eq:ratios}. The upper and lower lines show the band
due to 1-$\sigma$ uncertainty with respect to the central value as
defined in Eq.~\eqref{eq:dri}. \label{fig:m2-mlsp} }
\end{center}
\end{figure}

Before analyzing the expected accuracy of determination of stop
sector parameters let us study the possible influence of the
gaugino/higgsino sector parameters, taking as an example Scenario A.
The precise knowledge of the LSP mass and the mixing angles of the
charginos and the neutralinos may only be accessible after the
results from a linear $e^+e^-$ collider are available. In
Fig.~\ref{fig:m2-mlsp} we show the dependence of the normalized
ratios Eq.~\eqref{eq:ratios} on the gaugino mass parameter $M_2$ and
the mass of the LSP, $m_{\mathrm{LSP}} \equiv \mneu{1}$. In the left
plot of Fig.~\ref{fig:m2-mlsp} we keep the mass differences
$\mneu{2}-\mneu{1}$ and $\mcha{1}-\mneu{1} $ fixed as these are
expected to be measured with high precision at the LHC. As can be
seen, the value of $\widehat{R}_{1t}^{1b}$ is very stable in both
cases, whilst $\widehat{R}_{2t}^{1b}$ and $\widehat{R}_{2t}^{1t}$
exhibit an increase for larger values of $M_2$ and
$m_{\mathrm{LSP}}$. This is because both $\widehat{R}_{2t}^{1b}$ and
$\widehat{R}_{2t}^{1t}$ include a branching ratio for the decay to
$\neu{2}\, t$ that is close to its kinematic threshold. Therefore,
for an increasing $m_{\mathrm{LSP}}$ or $M_2$ (note that $M_1 \simeq
\mneu{1}$ and $M_2 \simeq \mneu{2}$ in Scenario~A) we approach the
point where this decay becomes impossible. High sensitivity of the
decay width near the threshold means that to use such a decay mode
to determine the mixing angle, one would have to know the masses
extremely precisely. In this case the ratio of branching ratios is
no longer a good observable. Moreover the branching ratio for such a
decay usually becomes very small.

\begin{figure}[t!]
\begin{center}
\includegraphics[width=0.47\textwidth]{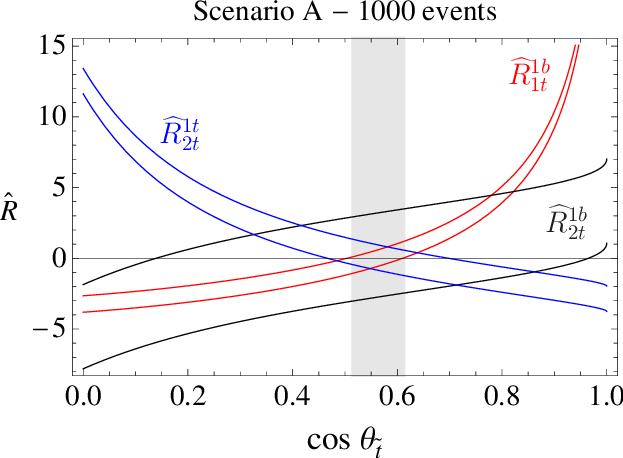}\hskip 0.5cm
\includegraphics[width=0.48\textwidth]{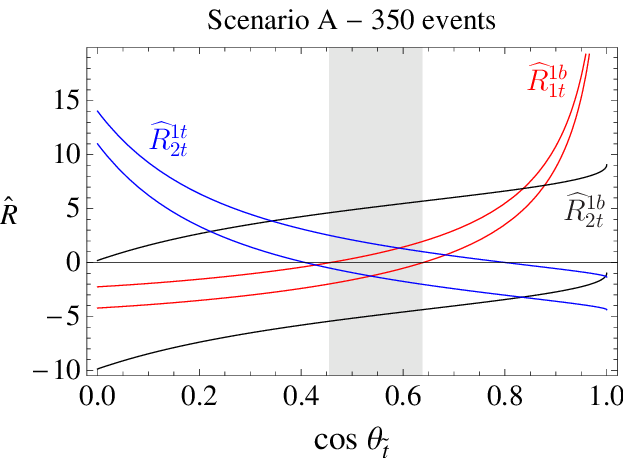}
\caption{Normalized ratios $\widehat{R}_j^i$,
Eq.~\eqref{eq:normalized}, near the nominal value of
$\cos\theta_{\st} = 0.56$ for Scenario~A. Red lines are for
$\widehat{R}_{1t}^{1b}$, black lines for $\widehat{R}_{2t}^{1b}$,
and blue lines for $\widehat{R}_{2t}^{1t}$, see
Eq.~\eqref{eq:ratios}. The upper and lower lines show the 1-$\sigma$
uncertainty as defined in the text. The region allowed by all three
ratios is shaded in grey. Left panel is for 1000 identified events,
right panel for 350 events. \label{fig:costdetermination}}
\end{center}
\end{figure}

In order to analyze the possible accuracy in extracting the mixing
parameters of the stop sector we start with the example of
Scenario~A. In Fig.~\ref{fig:costdetermination} we show the
behaviour of the normalized ratios of branching ratios,
Eq.~\eqref{eq:normalized}, near the nominal value of the mixing
angle $\cos \theta^{\mathrm{nominal}}_{\st}$ for 1000 and 350
events, left and right panel, respectively. Using only one of three
possible ratios, the smallest error and hence the best estimate we
get is using the ratio $\widehat{R}_{1t}^{1b}$, which depends on the
dominant decay modes $\neu{1}\, t $ and $\cha^+_1 b$. For the  ratio
$\widehat{R}_{2t}^{1t}$ the impact of the error is slightly larger
due to the limited statistics. On the other hand the ratio
$\widehat{R}_{2t}^{1b}$ gives the weakest constraints because both
$\cha^+_1$ and $\neu{2}$ are winos, hence their couplings to $\st_1$
follow a similar pattern. We assume here that the values of the
other SUSY parameters, including the $\st_1$ mass, are known. Using
only the information from the ratio $\widehat{R}_{1t}^{1b}$ we get
the estimate
\begin{eqnarray}
0.51 < \cos \theta_{\st} < 0.62
\end{eqnarray}
at the 1-$\sigma$ level, see Fig.~\ref{fig:costdetermination}, with
1000 events. We also show the plots when only 350 events are
included. In this case, we can expect the following sensitivity
\begin{eqnarray}
0.45 < \cos \theta_{\st} < 0.64\;,
\end{eqnarray}
with the allowed range increased by a factor $\sim 2$.

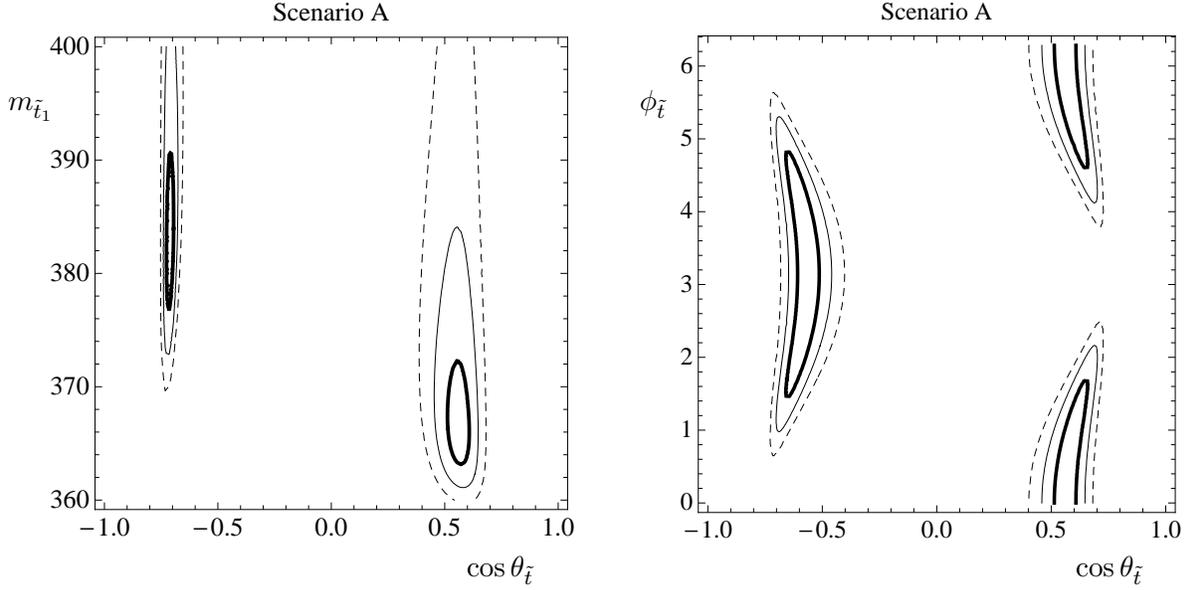
\begin{figure}
\begin{center}\setlength{\unitlength}{1cm}
\begin{picture}(12.5,7.8)
\put(-1.6,0.3){\mbox{\includegraphics[width=0.45\textwidth]{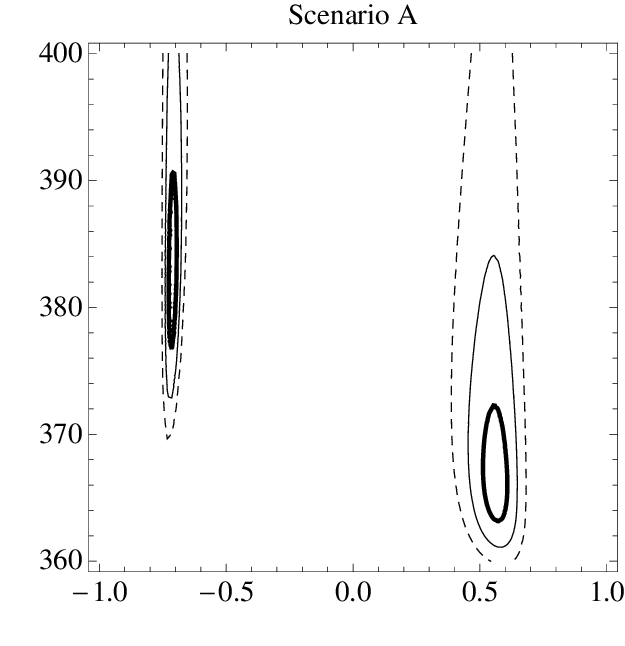}\hskip
0.5cm}} \put(-1.7,6.5){\small $m_{\st_1}$} \put(4.4,0.3){\small
$\cos \theta_{\st}$}
\put(6.8,.3){\includegraphics[width=0.43\textwidth]{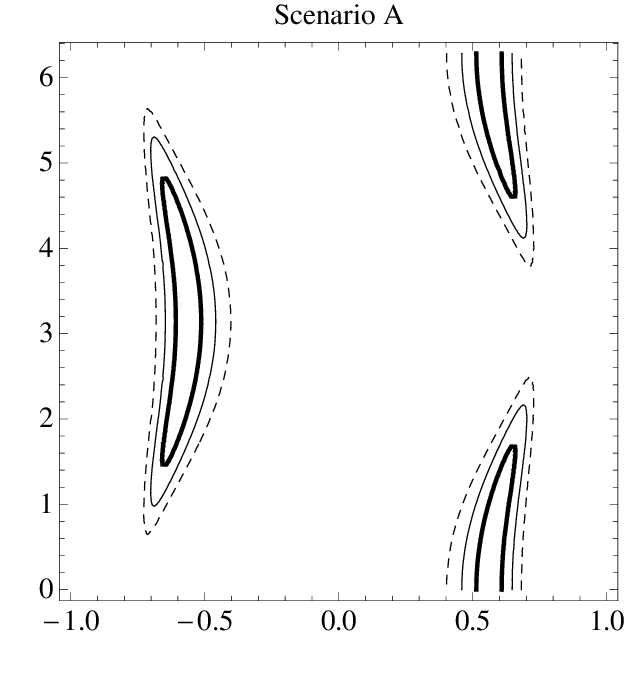}}
\put(6.7,6.5){\small $\phi_{\st}$} \put(12.5,0.3){\small $\cos
\theta_{\st}$}
\end{picture}
\caption{A $\chi^2$ fit of ratios ${R}_{1t}^{1b}$, ${R}_{2t}^{1b}$
and  ${R}_{2t}^{1t}$, Eq.~\eqref{eq:ratios}, to the stop mass and
the mixing angle $\cos\theta_{\st}$ (left panel), and the stop
mixing angle and the CP phase $\phi_{\st}$ (right panel) in Scenario
A for $n=1000$ events. Bold, normal and dashed lines are for 1-, 2-
and 3-$\sigma$ contours, respectively. \label{fig:chi2-scA}}
\end{center}
\end{figure}

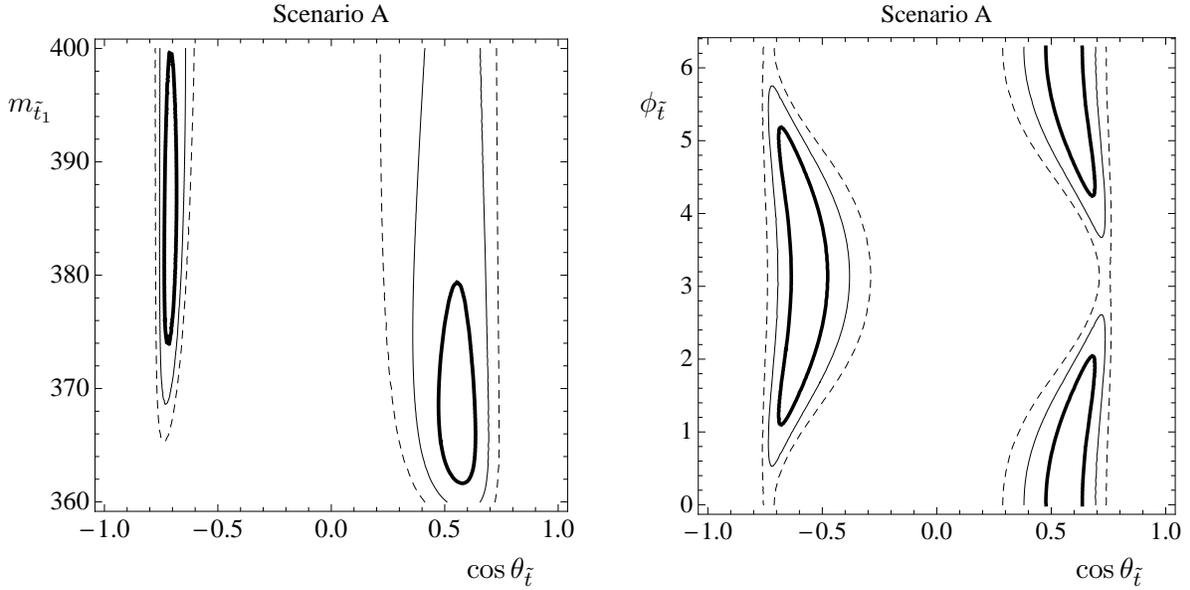
\begin{figure}
\begin{center}\setlength{\unitlength}{1cm}
\begin{picture}(12.5,7.8)
\put(-1.6,0.3){\mbox{\includegraphics[width=0.45\textwidth]{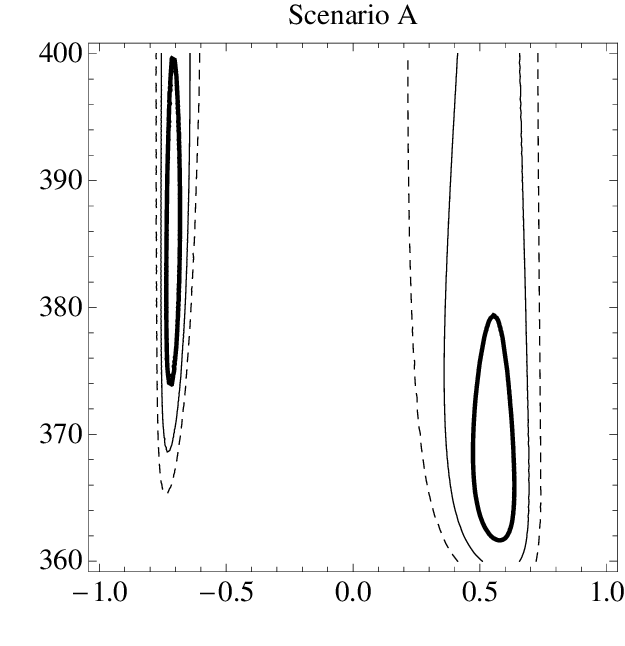}\hskip
0.5cm}} \put(-1.7,6.5){\small $m_{\st_1}$} \put(4.4,0.3){\small
$\cos \theta_{\st}$}
\put(6.8,.3){\includegraphics[width=0.43\textwidth]{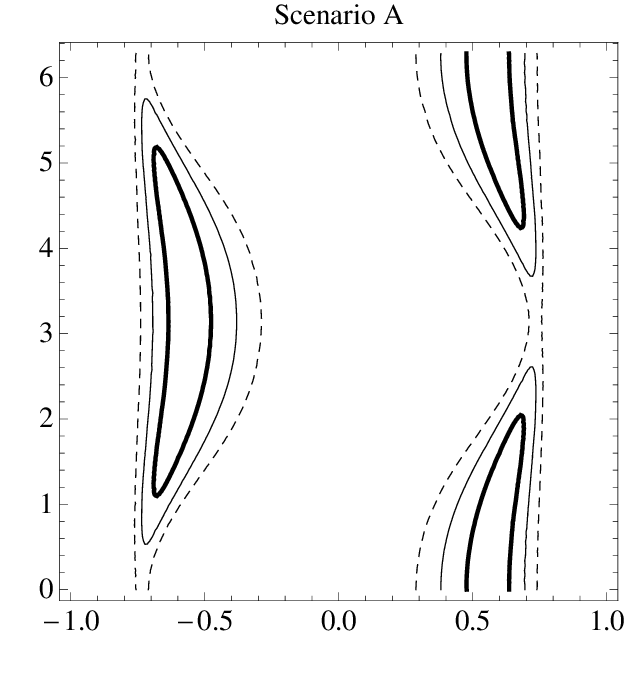}}
\put(6.7,6.5){\small $\phi_{\st}$} \put(12.5,0.3){\small $\cos
\theta_{\st}$}
\end{picture}
\caption{Same as Fig.~\ref{fig:chi2-scA}, Scenario~A, but for 350
identified stop decays. \label{fig:chi2-scA-350}}
\end{center}
\end{figure}

Having at hand three possible decay modes we can constrain not only
the mixing angle $\cos\theta_{\st}$ but also the mass of the light
stop quark and the CP-violating phase $\phi_{\st}$. We use a
$\chi^2$ fit defined as follows:
\begin{eqnarray}
\chi^2 = \sum_{\{i,j\}} \left(\frac{R_j^i(\cos\theta_{\st}) -
R_j^{i\:\mathrm{nominal}}}{\Delta R_j^i}\right)^2\,,
\end{eqnarray}
where the error is defined by Eq.~\eqref{eq:dri} and the sum runs
over the respective ratios for each of the scenarios, e.g.\ $\{i,j\}
= \{\{1b,1t\},\{1b,2t\},\{1t,2t\}\}$ in Scenario A, cf.\
Eqs~\eqref{eq:ratios}--\eqref{eq:ratiosC1}. The results of fitting
the stop mass $m_{\st_1}$ and the mixing angle $\cos\theta_{\st}$ in
Scenario A are shown in the left panel of Fig.~\ref{fig:chi2-scA}.
We find two minima of $\chi^2$ that fit the input data well. In
order to resolve the two-fold ambiguity, additional observables will
be needed. Assuming that we can pin down the correct solution we get
the following 1-$\sigma$ (2-$\sigma$) estimate of the two parameters
\begin{eqnarray}
& & m_{\st_1} = 366^{+6\; (+18)}_{-3\; (-5)}\,, \qquad
\cos\theta_{\st} = 0.56 \pm 0.05\; (\pm 0.1)\,, \qquad \theta_{\st}
= 0.98^{+0.06\; (+0.11)}_{-0.06\;(-0.13)}
\end{eqnarray}
for 1000 events assuming the errors defined in Eqs.~\eqref{eq:error}
and \eqref{eq:dri} and neglecting other uncertainties. The mixing
angle obtained here should be treated as an effective parameter, as
discussed in Sec.~\ref{sec:formulae}. The better lower bound for the
measured mass is a consequence of reaching the kinematic threshold
for the decay ${\st_1}\to \neu{2} t$. In the right panel of
Fig.~\ref{fig:chi2-scA} we show the results of the $\chi^2$ fit to
the mixing angle and the phase $\phi_{\st}$. As expected, the
sensitivity to the CP phase is poor and taking into account the
possible ambiguity in the mixing angle $\cos\theta_{\st}$, the full
range of phases remains allowed. The same set of plots in the case
of 350 identified events is shown in Fig.~\ref{fig:chi2-scA-350},
giving
\begin{eqnarray}
& & m_{\st_1} = 366^{+14\; (+ 80)}_{-3\;(-7)}\,, \qquad
\cos\theta_{\st} = 0.56 \pm 0.08\; (\pm 0.15)\,, \qquad \theta_{\st}
= 0.98^{+0.10\; (+0.17)}_{-0.10\;(-0.20)}\;,
\end{eqnarray}
at the 1-$\sigma$ (2-$\sigma$) level, respectively.

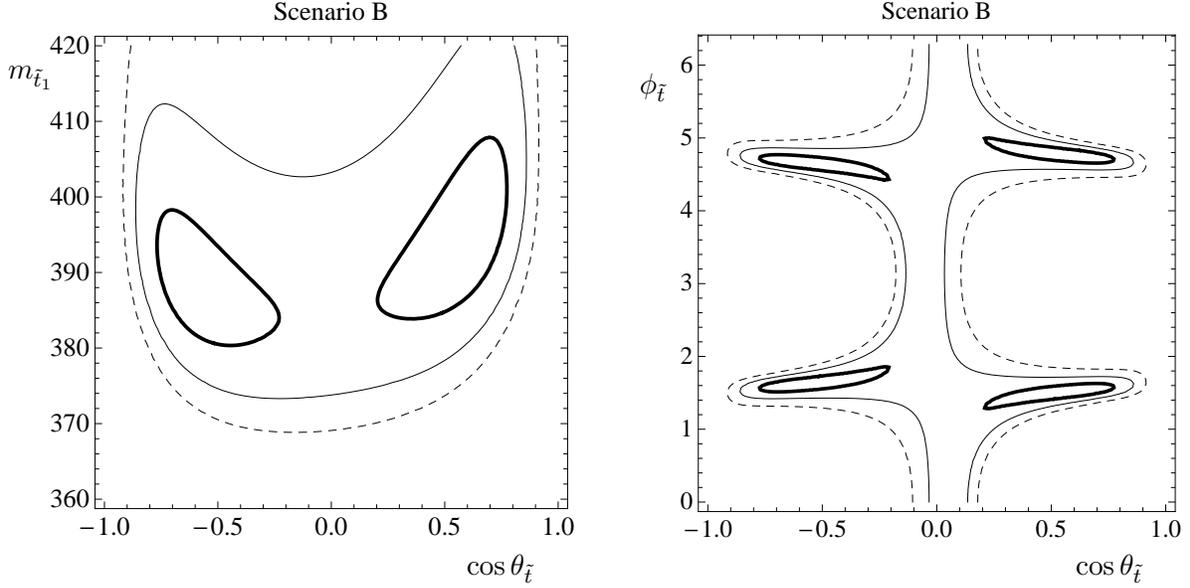
\begin{figure}
\begin{center}\setlength{\unitlength}{1cm}
\begin{picture}(12.5,7.8)
\put(-1.6,0.3){\mbox{\includegraphics[width=0.45\textwidth]{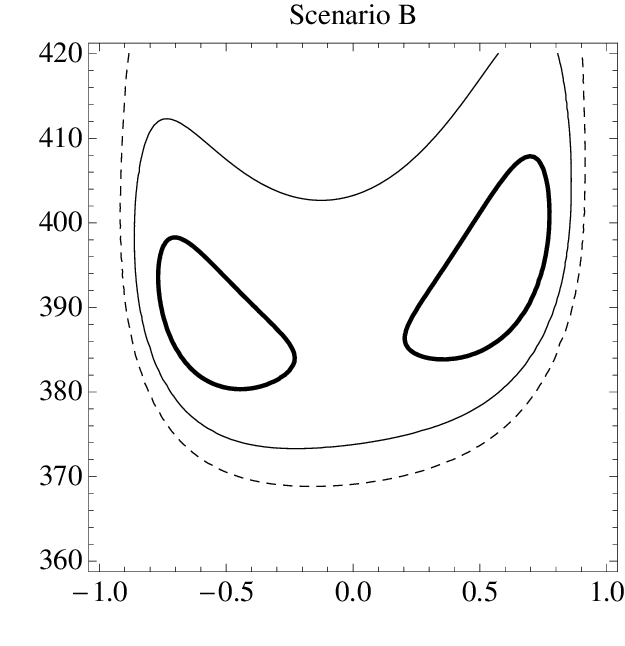}\hskip
0.5cm}} \put(-1.7,6.9){\small $m_{\st_1}$} \put(4.4,0.3){\small
$\cos \theta_{\st}$}
\put(6.8,.3){\includegraphics[width=0.43\textwidth]{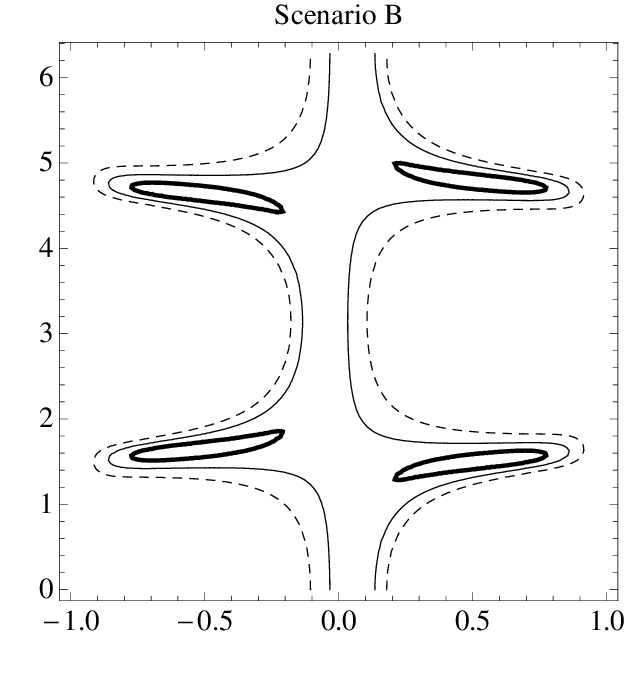}}
\put(6.7,6.7){\small $\phi_{\st}$} \put(12.5,0.3){\small $\cos
\theta_{\st}$}
\end{picture}
\caption{A $\chi^2$ fit of ratios ${R}_{1t}^{1b}$, ${R}_{2t}^{1b}$,
${R}_{2t}^{1t}$, ${R}_{2b}^{1b}$, ${R}_{2b}^{1t}$ and
${R}_{2b}^{2t}$, Eq.~\eqref{eq:ratios} and Eq.~\eqref{eq:ratiosB},
to the stop mass and the mixing angle $\cos\theta_{\st}$ (left
panel), and the stop mixing angle and the CP phase $\phi_{\st}$
(right panel) in Scenario B for $n=1000$ events. Bold, normal and
dashed lines are for 1-, 2- and 3-$\sigma$ contours, respectively.
\label{fig:chi2-scB}}
\end{center}
\end{figure}

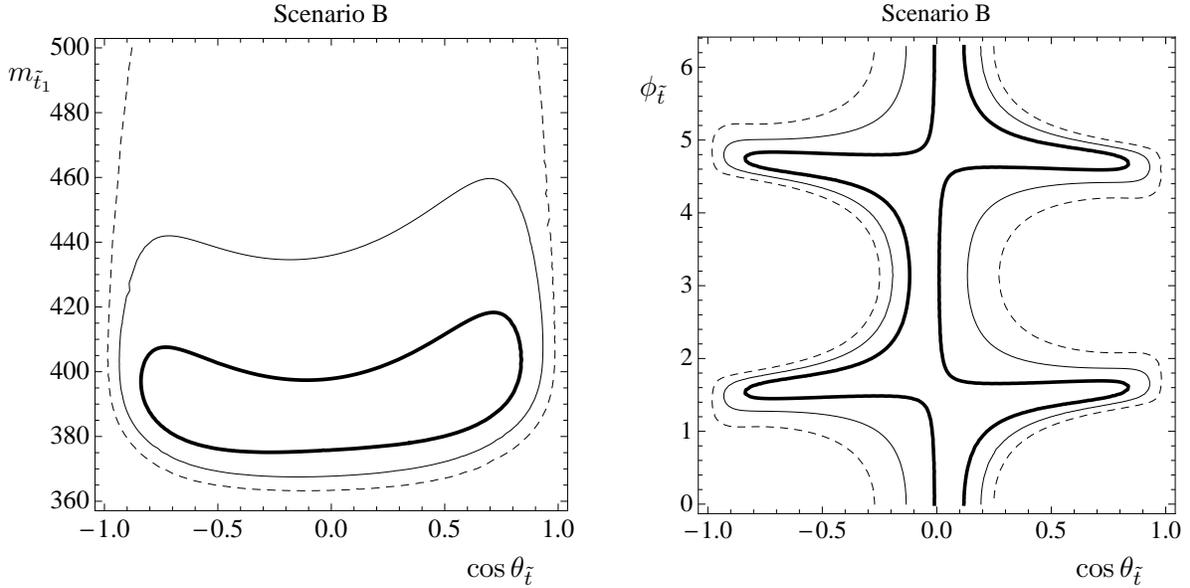
\begin{figure}[!t]
\begin{center}\setlength{\unitlength}{1cm}
\begin{picture}(12.5,7.8)
\put(-1.6,0.3){\mbox{\includegraphics[width=0.45\textwidth]{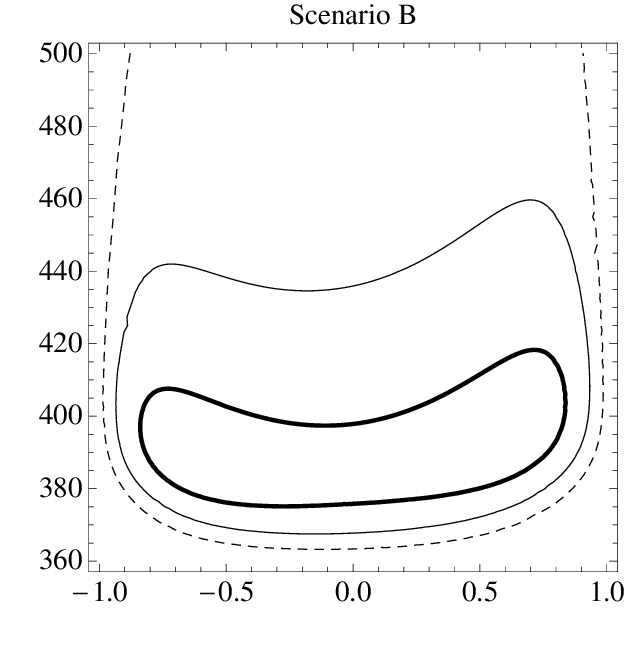}\hskip
0.5cm}} \put(-1.7,6.9){\small $m_{\st_1}$} \put(4.4,0.3){\small
$\cos \theta_{\st}$}
\put(6.8,.3){\includegraphics[width=0.43\textwidth]{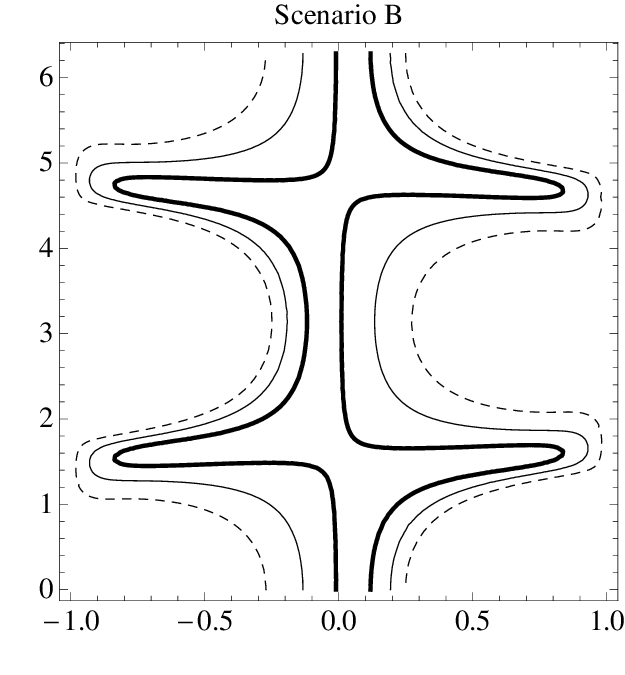}}
\put(6.7,6.7){\small $\phi_{\st}$} \put(12.5,0.3){\small $\cos
\theta_{\st}$}
\end{picture}
\caption{Same as Fig.~\ref{fig:chi2-scB}, Scenario~B, but for 350
identified stop decays. \label{fig:chi2-scB-350}}
\end{center}
\end{figure}

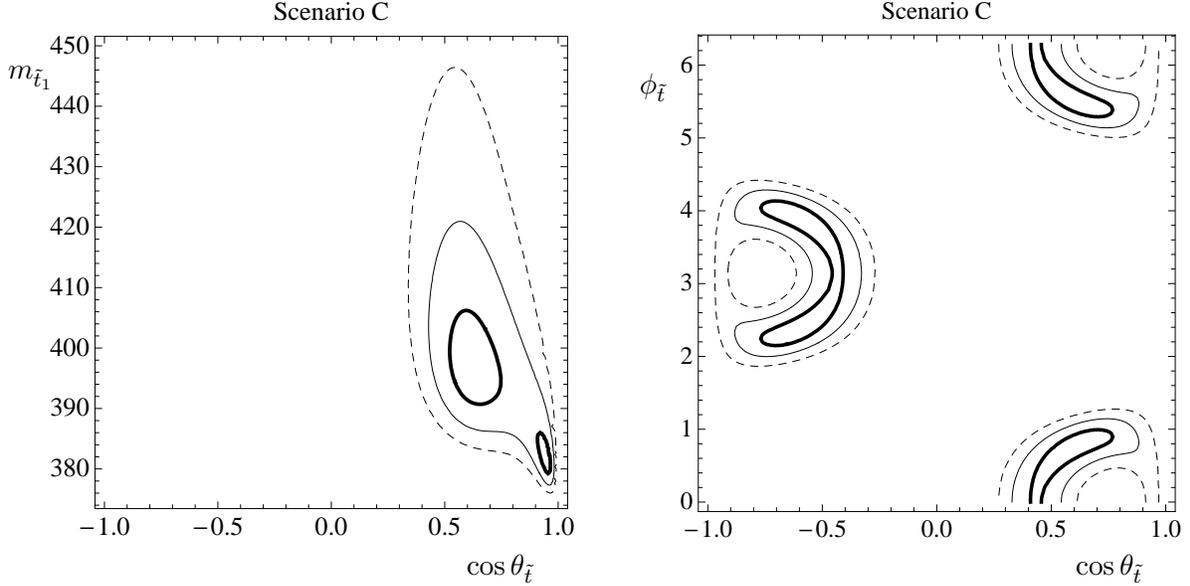
\begin{figure}[t!]
\begin{center}\setlength{\unitlength}{1cm}
\begin{picture}(12.5,7.8)
\put(-1.6,0.3){\mbox{\includegraphics[width=0.45\textwidth]{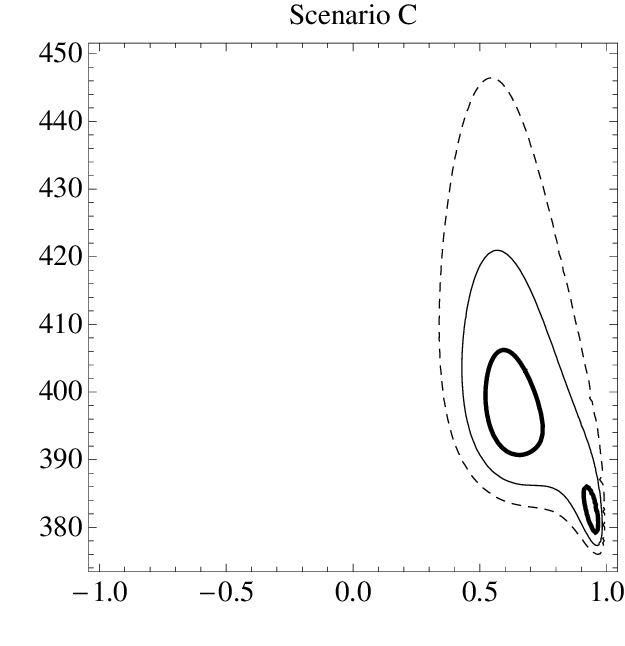}\hskip
0.5cm}} \put(-1.7,6.9){\small $m_{\st_1}$} \put(4.4,0.3){\small
$\cos \theta_{\st}$}
\put(6.8,.3){\includegraphics[width=0.43\textwidth]{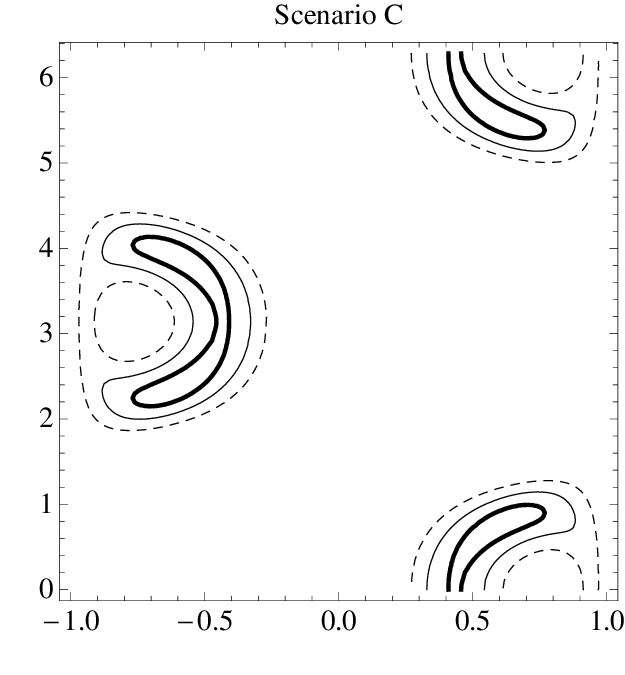}}
\put(6.7,6.7){\small $\phi_{\st}$} \put(12.5,0.3){\small $\cos
\theta_{\st}$}
\end{picture}
\caption{A $\chi^2$ fit of ratios ${R}_{1t}^{1b}$, ${R}_{2t}^{1b}$,
${R}_{2t}^{1t}$, ${R}_{3t}^{1b}$, ${R}_{3t}^{1t}$ and
${R}_{3t}^{2t}$, Eq.~\eqref{eq:ratios} and Eq.~\eqref{eq:ratiosC},
to the stop mass and the mixing angle $\cos\theta_{\st}$ (left
panel) and the stop mixing angle and the CP phase $\phi_{\st}$
(right panel) in Scenario C for $n=1000$ events. Bold, normal and
dashed lines are for 1-, 2- and 3-$\sigma$ contours, respectively.
\label{fig:chi2-scC1}}
\end{center}
\end{figure}

\begin{figure}[t!]
\begin{center}\setlength{\unitlength}{1cm}
\begin{picture}(12.5,7.8)
\put(-1.6,0.3){\mbox{\includegraphics[width=0.45\textwidth]{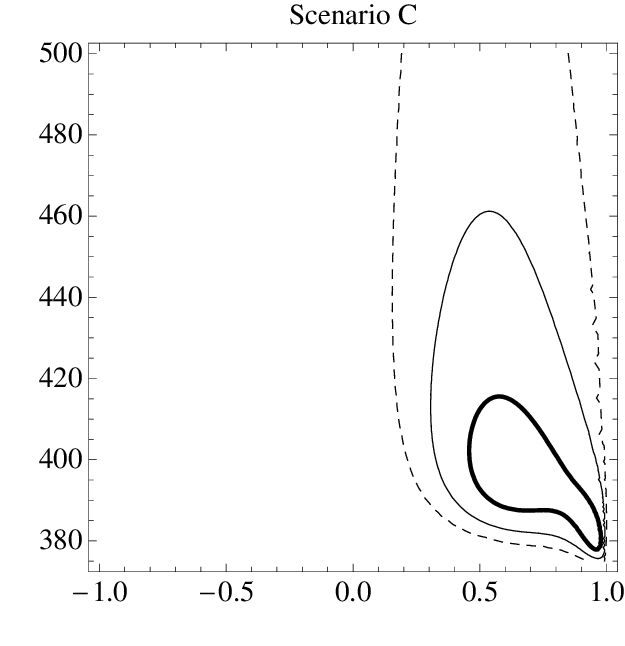}\hskip
0.5cm}} \put(-1.7,6.9){\small $m_{\st_1}$} \put(4.4,0.3){\small
$\cos \theta_{\st}$}
\put(6.8,.3){\includegraphics[width=0.43\textwidth]{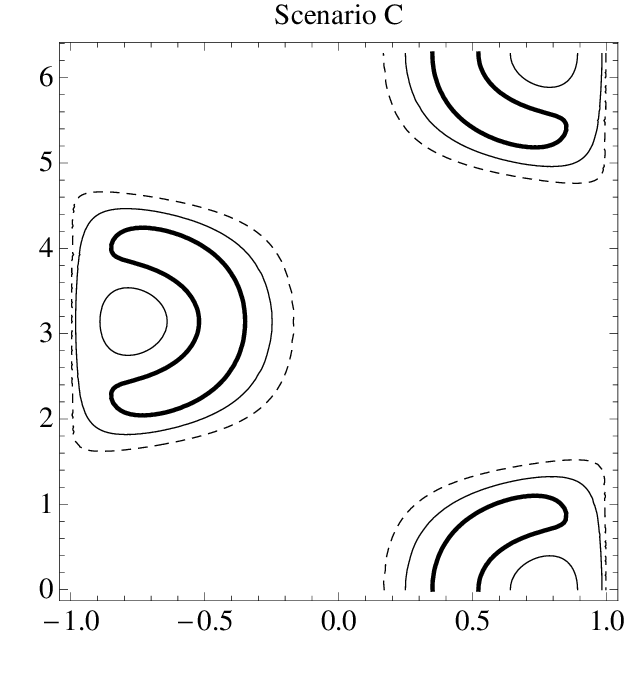}}
\put(6.7,6.7){\small $\phi_{\st}$} \put(12.5,0.3){\small $\cos
\theta_{\st}$}
\end{picture}
\caption{Same as Fig.~\ref{fig:chi2-scC1}, Scenario~C, but for 350
identified stop decays. \label{fig:chi2-scC1-350}}
\end{center}
\end{figure}
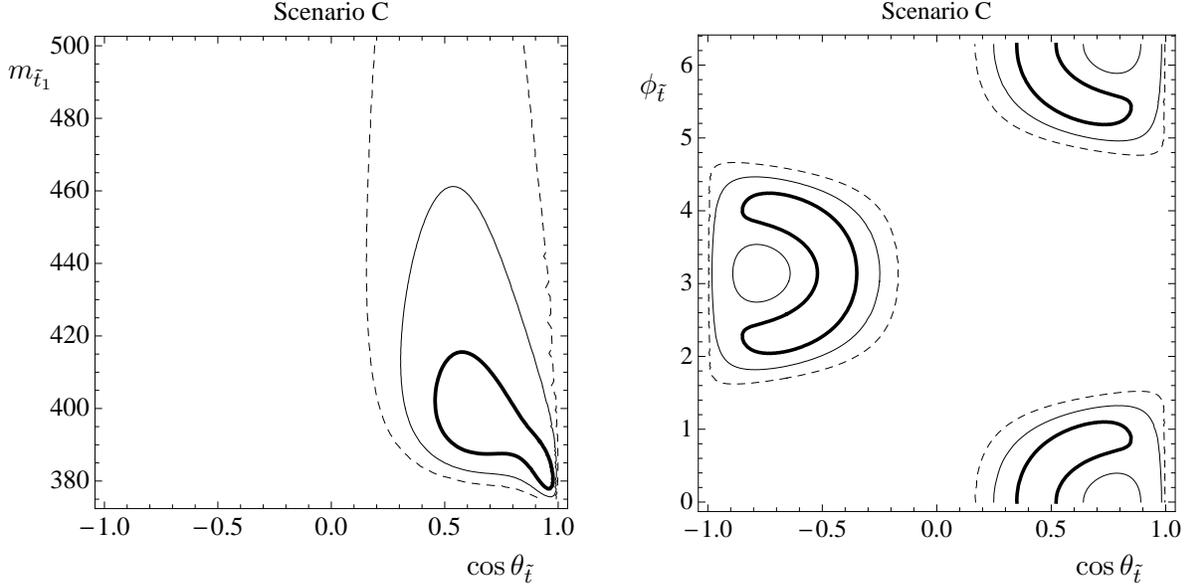

The situation changes for Scenarios B and C. We now have 6 possible
ratios in each case, for Scenario B: ${R}_{1t}^{1b}$,
${R}_{2t}^{1b}$, ${R}_{2t}^{1t}$, ${R}_{2b}^{1b}$, ${R}_{2b}^{1t}$,
${R}_{2b}^{2t}$, and for Scenario C: ${R}_{1t}^{1b}$,
${R}_{2t}^{1b}$, ${R}_{2t}^{1t}$, ${R}_{3t}^{1b}$, ${R}_{3t}^{1t}$,
${R}_{3t}^{2t}$. The results of the fit in Scenario~B have been
shown in Figs.~\ref{fig:chi2-scB} and~\ref{fig:chi2-scB-350} for
1000 and 350 events, respectively. Corresponding plots for
Scenario~C are presented in Figs.~\ref{fig:chi2-scC1}
and~\ref{fig:chi2-scC1-350}. We again consider two cases: fitting of
the mass $m_{\st}$ together with the mixing angle $\cos\theta_{\st}$
and fitting of the mixing angle together with the CP-violating
phase. In both cases we assume that the value of the third parameter
is known. Charginos and neutralinos now have a significant higgsino
component and, as we saw in Figs.~\ref{fig:scB-gamma}
and~\ref{fig:scC-gamma}, the dependence on the mixing angle is much
weaker. Therefore the constraints for the mixing angle and the mass
that we get are not as good as in the case of Scenario A. It is
interesting to note that in general the results of the fit are
better in Scenarios A and C (gaugino and higgsino, respectively)
than in Scenario B (mixed case). Consequently we conclude that the
scenario with strong mixing between gauginos and higgsinos would be
the most difficult to resolve. This is visible for the 350-event
case in Scenario~B (Fig.~\ref{fig:chi2-scB-350}), where the mixing
angle and the phase are practically unconstrained.

Analyzing both the mixing angle and the phase, we obtain four
allowed regions. Nevertheless smaller regions are allowed for the CP
phase as our observables are more sensitive to it than in Scenario
A. Branching ratios are CP-even observables, therefore they cannot
resolve ambiguities for the CP phase. This shows that for precise
measurements in the stop sector one has to use CP-odd observables,
like triple products of
momenta~\cite{MoortgatPick:2010wp,Ellis:2008hq,Deppisch:2009nj}.
Only such a combined analysis of CP-even and CP-odd observables can
give an unambiguous determination of the stop sector parameters.

Finally, we note that in Scenario C it might be difficult to resolve
the decays $\neu{2}\, t$ and $\neu{3}\, t$, as the two close-in-mass
higgsino-like neutralinos may be difficult to disentangle at the
LHC. After combining decay modes of the two neutralinos we lose much
of the sensitivity to the elements of the stop mixing matrix. In
such cases, precise measurements in the neutralino and chargino
sectors will be essential. The measurements of the masses and
couplings of these particles may be possible with high accuracy at
the linear collider.

\section{Conclusions}\label{sec:conclusions}

As stops play an important role in the MSSM, it is crucial to
measure their couplings and masses at future colliders in order to
understand the underlying model. Therefore we have proposed a
promising way to get hints of the stop sector parameters at the LHC
by studying the dependence of branching ratios in various mixing
scenarios. In particular, we have discussed the couplings and the
decays of the supersymmetric top partners to the charginos and the
neutralinos.

A careful analysis of the couplings of scalar tops to electroweak
gauginos and higgsinos shows a strong dependence on the mixing angle
and the CP-violating phase of the stop sector. This effect arises
due to the structure of the electroweak gauge couplings and the
Yukawa couplings of left and right stop states. We have analyzed
three benchmark scenarios with different structures for the gaugino
and higgsino sectors, where the light charginos and neutralinos had
gaugino-like, higgsino-like or mixed composition. Analysis of the
decay widths and the branching ratios has shown a strong relation
between the stop mixing parameters and the decay pattern in each of
the scenarios.

Next, we have discussed a possible approach to get a handle on the
light stop mass, the mixing angle $\theta_{\st}$ and the
CP-violating phase $\phi_{\st}$ at the Large Hadron Collider. Since
stops will be produced in large numbers at this machine one can hope
to learn the stop properties from their decay pattern. As the
branching ratios are going to be difficult to be measured at the
LHC, we propose to analyze the ratios of branching ratios for
different decay modes. These observables inherit a strong dependence
on the mixing angle from stop decay widths and therefore can be a
sensitive probe of the stop sector. Since they rely only on the
relative numbers of stops decaying via various channels, many
experimental uncertainties will cancel. In particular, one does not
need to control all of the possible decay modes. In fact, as we have
shown for the SPS1a$'$ parameter point, using only two decay modes
can give good constraints on the stop mixing angle. Finally we have
performed $\chi^2$ fits to show that the ratios of branching ratios
can give strong bounds on the parameters of the stop sector: the
mass of $\st_1$, the stop mixing angle $\cos\theta_{\st}$ and the
CP-violating phase $\phi_{\st}$. The expected accuracy depends upon
the scenario studied but looks the most promising for mSUGRA models.

In the present study many of the experimental uncertainties have
been neglected. Backgrounds have not been explicitly included, but
assuming that they can be successfully controlled, as discussed, we
extrapolated their possible impact by doubling the usual statistical
error. It is clear that more detailed experimental studies are
needed to assess the feasibility of the method in the harsh LHC
environment and its possible accuracy. However, taking into account
the importance of the stop sector for our understanding of the
supersymmetric model, we think that such a study deserves further
attention. Application of this method will require the study of many
possible final states to understand those that are most promising.
In addition a good control of detector effects, like fake rates for
leptons and $b$-jets, and SM as well as SUSY backgrounds will be
needed, but these are not yet included. These uncertainties will
certainly lead to the method presented here having lower
sensitivity. However, we believe that the results obtained here,
even with a rather low number of events, are encouraging enough to
pursue further studies of precision measurements in the stop sector.

\section*{Acknowledgements}
We want to thank Philip Bechtle, Klaus Desch, Jan Kalinowski, Filip
Moortgat and Peter Wienemann for interesting discussions. KR is
supported by the EU Network MRTN-CT-2006-035505 ``Tools and
Precision Calculations for Physics Discoveries at Colliders''
(HEPTools). JT is supported by the UK Science and Technology
Facilities Council (STFC).



\begin{thebibliography}{19}

\bibitem{susy}
  Yu.~A.~Golfand and E.~P.~Likhtman,
  JETP Lett.\  {\bf 13} (1971) 323
  [Pisma Zh.\ Eksp.\ Teor.\ Fiz.\  {\bf 13} (1971) 452];
  D.~V.~Volkov and V.~P.~Akulov,
  Phys.\ Lett.\  B {\bf 46} (1973) 109;
  J.~Wess and B.~Zumino,
  Phys.\ Lett.\  B {\bf 49} (1974) 52;
  Nucl.\ Phys.\  B {\bf 70} (1974) 39;
  for reviews see e.g.
  H.~P.~Nilles, Phys. Rept. {\bf 110} (1984)  1;
  H.~E.~Haber and G.~L.~Kane,
  Phys.\ Rept.\  {\bf 117} (1985) 75;
  D.~J.~H.~Chung, L.~L.~Everett, G.~L.~Kane, S.~F.~King, J.~D.~Lykken and L.~T.~Wang,
  Phys.\ Rept.\  {\bf 407} (2005) 1
  [arXiv:hep-ph/0312378].

\bibitem{Ellis:2007fu}
  J.~R.~Ellis, S.~Heinemeyer, K.~A.~Olive, A.~M.~Weber and G.~Weiglein,
  JHEP {\bf 0708} (2007) 083
  [arXiv:0706.0652 [hep-ph]];
  O.~Buchmueller {\it et al.},
  JHEP {\bf 0809} (2008) 117
  [arXiv:0808.4128 [hep-ph]].

\bibitem{Ball:2007zza}
  G.~L.~Bayatian {\it et al.}  [CMS Collaboration],
  J.\ Phys.\ G {\bf 34} (2007) 995.

\bibitem{Aad:2009wy}
  G.~Aad {\it et al.}  [The ATLAS Collaboration],
  arXiv:0901.0512 [hep-ex].

\bibitem{Hinchliffe:1996iu}
  I.~Hinchliffe, F.~E.~Paige, M.~D.~Shapiro, J.~Soderqvist and W.~Yao,
  Phys.\ Rev.\  D {\bf 55} (1997) 5520
  [arXiv:hep-ph/9610544].

\bibitem{AguilarSaavedra:2005pw}
  J.~A.~Aguilar-Saavedra {\it et al.},
  Eur.\ Phys.\ J.\  C {\bf 46} (2006) 43
  [arXiv:hep-ph/0511344].

\bibitem{Hisano:2003qu}
  J.~Hisano, K.~Kawagoe, R.~Kitano and M.~M.~Nojiri,
  Phys.\ Rev.\  D {\bf 66} (2002) 115004
  [arXiv:hep-ph/0204078];
  J.~Hisano, K.~Kawagoe and M.~M.~Nojiri,
  Phys.\ Rev.\  D {\bf 68} (2003) 035007
  [arXiv:hep-ph/0304214].

\bibitem{Carena:2008vj}
  M.~Carena, G.~Nardini, M.~Quiros and C.~E.~M.~Wagner,
  Nucl.\ Phys.\  B {\bf 812} (2009) 243
  [arXiv:0809.3760 [hep-ph]].

\bibitem{Ellis:1990nz}
  J.~R.~Ellis, G.~Ridolfi and F.~Zwirner,
  Phys.\ Lett.\  B {\bf 257} (1991) 83;
  M.~Frank, T.~Hahn, S.~Heinemeyer, W.~Hollik, H.~Rzehak and G.~Weiglein,
  JHEP {\bf 0702} (2007) 047
  [arXiv:hep-ph/0611326].

\bibitem{Pilaftsis:1998pe}
  A.~Pilaftsis,
  Phys.\ Rev.\  D {\bf 58} (1998) 096010
  [arXiv:hep-ph/9803297];
  A.~Pilaftsis,
  Phys.\ Lett.\  B {\bf 435} (1998) 88
  [arXiv:hep-ph/9805373].

\bibitem{Bartl:2000kw}
  M.~Berggren, R.~Keranen, H.~Kluge and A.~Sopczak,
  arXiv:hep-ph/9911345;
  A.~Bartl, H.~Eberl, S.~Kraml, W.~Majerotto and W.~Porod,
  Eur.\ Phys.\ J.\ direct C {\bf 2} (2000) 6
  [arXiv:hep-ph/0002115];
  R.~Keranen, A.~Sopczak, H.~Kluge and M.~Berggren,
  Eur.\ Phys.\ J.\ direct C {\bf 2} (2000) 7;
  E.~Boos, H.~U.~Martyn, G.~A.~Moortgat-Pick, M.~Sachwitz, A.~Sherstnev and P.~M.~Zerwas,
  Eur.\ Phys.\ J.\  C {\bf 30} (2003) 395
  [arXiv:hep-ph/0303110];
  A.~Arhrib and W.~Hollik,
  JHEP {\bf 0404} (2004) 073
  [arXiv:hep-ph/0311149].

\bibitem{Freitas:2007zr}
  A.~Freitas, C.~Milstene, M.~Schmitt and A.~Sopczak,
  JHEP {\bf 0809} (2008) 076
  [arXiv:0712.4010 [hep-ph]].

\bibitem{Kovarik:2004vd}
  K.~Kovarik, C.~Weber, H.~Eberl and W.~Majerotto,
  Phys.\ Lett.\  B {\bf 591} (2004) 242
  [arXiv:hep-ph/0401092];
  K.~Kovarik, C.~Weber, H.~Eberl and W.~Majerotto,
  Phys.\ Rev.\  D {\bf 72} (2005) 053010
  [arXiv:hep-ph/0506021].

\bibitem{Bartl:1996wt}
  A.~Bartl, H.~Eberl, S.~Kraml, W.~Majerotto and W.~Porod,
  Z.\ Phys.\  C {\bf 73} (1997) 469
  [arXiv:hep-ph/9603410];
  A.~Bartl, H.~Eberl, S.~Kraml, W.~Majerotto, W.~Porod and A.~Sopczak,
  Z.\ Phys.\  C {\bf 76} (1997) 549
  [arXiv:hep-ph/9701336].

\bibitem{kramlphd}
  S.~Kraml, Ph.D. thesis, Vienna 1999,
  arXiv:hep-ph/9903257.

\bibitem{finch}
  A.~Finch, H.~Nowak and A.~Sopczak,
  arXiv:hep-ph/0211140.

\bibitem{Baer:1991cb}
  H.~Baer, M.~Drees, R.~Godbole, J.~F.~Gunion and X.~Tata,
  Phys.\ Rev.\  D {\bf 44} (1991) 725;
  H.~Baer, J.~Sender and X.~Tata,
  Phys.\ Rev.\  D {\bf 50} (1994) 4517
  [arXiv:hep-ph/9404342].

\bibitem{beenakker98}
  W.~Beenakker, M.~Kramer, T.~Plehn, M.~Spira and P.~M.~Zerwas,
  Nucl.\ Phys.\  B {\bf 515} (1998) 3
  [arXiv:hep-ph/9710451].

\bibitem{Weiglein:2004hn}
  G.~Weiglein {\it et al.}  [LHC/LC Study Group],
  Phys.\ Rept.\  {\bf 426} (2006) 47
  [arXiv:hep-ph/0410364].

\bibitem{Casadei:2010nf}
  D.~Casadei, R.~Konoplich and R.~Djilkibaev,
  Phys.\ Rev.\  D {\bf 82} (2010) 075011
  [arXiv:1006.5875 [hep-ph]].

\bibitem{Perelstein:2008zt}
  M.~Perelstein and A.~Weiler,
  JHEP {\bf 0903}, 141 (2009)
  [arXiv:0811.1024 [hep-ph]].

\bibitem{Meade:2006dw}
  P.~Meade and M.~Reece,
  Phys.\ Rev.\  D {\bf 74} (2006) 015010
  [arXiv:hep-ph/0601124].

\bibitem{Bechtle:2009ty}
  P.~Bechtle, K.~Desch and P.~Wienemann,
  Comput.\ Phys.\ Commun.\  {\bf 174} (2006) 47
  [arXiv:hep-ph/0412012];
  P.~Bechtle, K.~Desch, M.~Uhlenbrock and P.~Wienemann,
  Eur.\ Phys.\ J.\  C {\bf 66} (2010) 215
  [arXiv:0907.2589 [hep-ph]].

\bibitem{Lafaye:2007vs}
  R.~Lafaye, T.~Plehn, M.~Rauch and D.~Zerwas,
  Eur.\ Phys.\ J.\  C {\bf 54} (2008) 617
  [arXiv:0709.3985 [hep-ph]].

\bibitem{Bartl:1994bu}
  A.~Bartl, W.~Majerotto and W.~Porod,
  Z.\ Phys.\  C {\bf 64} (1994) 499
  [Erratum-ibid.\  C {\bf 68} (1995) 518].

\bibitem{Hidaka:2000cm}
  A.~Bartl {\it et al.},
  Phys.\ Lett.\  B {\bf 435} (1998) 118
  [arXiv:hep-ph/9804265];
  K.~Hidaka and A.~Bartl,
  Phys.\ Lett.\  B {\bf 501} (2001) 78
  [arXiv:hep-ph/0012021].

\bibitem{Bartl:2003pd}
  A.~Bartl, S.~Hesselbach, K.~Hidaka, T.~Kernreiter and W.~Porod,
  Phys.\ Rev.\  D {\bf 70} (2004) 035003
  [arXiv:hep-ph/0311338].

\bibitem{nlo1}
  S.~Kraml, H.~Eberl, A.~Bartl, W.~Majerotto and W.~Porod,
  Phys.\ Lett.\  B {\bf 386} (1996) 175
  [arXiv:hep-ph/9605412].

\bibitem{nlo2}
  A.~Djouadi, W.~Hollik and C.~Junger,
  Phys.\ Rev.\  D {\bf 55} (1997) 6975
  [arXiv:hep-ph/9609419].

\bibitem{nlo3}
  W.~Beenakker, R.~Hopker, T.~Plehn and P.~M.~Zerwas,
  Z.\ Phys.\  C {\bf 75} (1997) 349
  [arXiv:hep-ph/9610313].

\bibitem{Bartl:1997pb}
  A.~Bartl, H.~Eberl, K.~Hidaka, S.~Kraml, W.~Majerotto, W.~Porod and Y.~Yamada,
  bosons,''
  Phys.\ Lett.\  B {\bf 419} (1998) 243
  [arXiv:hep-ph/9710286];
  A.~Bartl, H.~Eberl, K.~Hidaka, S.~Kraml, W.~Majerotto, W.~Porod and Y.~Yamada,
  Phys.\ Rev.\  D {\bf 59} (1999) 115007
  [arXiv:hep-ph/9806299].

\bibitem{Guasch:2001kz}
  J.~Guasch, W.~Hollik and J.~Sola,
  Phys.\ Lett.\  B {\bf 437} (1998) 88
  [arXiv:hep-ph/9802329];
  Phys.\ Lett.\  B {\bf 510} (2001) 211
  [arXiv:hep-ph/0101086].

\bibitem{full1loop}
  J.~Guasch, W.~Hollik and J.~Sola,
  JHEP {\bf 0210} (2002) 040
  [arXiv:hep-ph/0207364].

\bibitem{nlo4}
  A.~Arhrib and R.~Benbrik,
  Phys.\ Rev.\  D {\bf 71} (2005) 095001
  [arXiv:hep-ph/0412349].

\bibitem{Plehn:2010st}
  T.~Plehn, M.~Spannowsky, M.~Takeuchi and D.~Zerwas,
  JHEP {\bf 1010} (2010) 078
  [arXiv:1006.2833 [hep-ph]].

\bibitem{MoortgatPick:2010wp}
  G.~Moortgat-Pick, K.~Rolbiecki and J.~Tattersall,
  arXiv:1008.2206 [hep-ph].

\bibitem{Plehn:2009rk}
  J.~M.~Butterworth, A.~R.~Davison, M.~Rubin and G.~P.~Salam,
  Phys.\ Rev.\ Lett.\  {\bf 100} (2008) 242001
  [arXiv:0802.2470 [hep-ph]];
  T.~Plehn, G.~P.~Salam and M.~Spannowsky,
  Phys.\ Rev.\ Lett.\  {\bf 104} (2010) 111801
  [arXiv:0910.5472 [hep-ph]];
  D.~E.~Soper and M.~Spannowsky,
  JHEP {\bf 1008} (2010) 029
  [arXiv:1005.0417 [hep-ph]].

\bibitem{Desch:2003vw}
  K.~Desch, J.~Kalinowski, G.~A.~Moortgat-Pick, M.~M.~Nojiri and G.~Polesello,
  JHEP {\bf 0402} (2004) 035
  [arXiv:hep-ph/0312069].

\bibitem{Rosiek:1989rs}
  J.~Rosiek,
  Phys.\ Rev.\  D {\bf 41} (1990) 3464.

\bibitem{Hollik:2003jj}
  W.~Hollik and H.~Rzehak,
  Eur.\ Phys.\ J.\  C {\bf 32} (2003) 127
  [arXiv:hep-ph/0305328];
  S.~Heinemeyer, H.~Rzehak and C.~Schappacher,
  Phys.\ Rev.\  D {\bf 82} (2010) 075010
  [arXiv:1007.0689 [hep-ph]].

\bibitem{Ellis:2008hq}
  J.~Ellis, F.~Moortgat, G.~Moortgat-Pick, J.~M.~Smillie and J.~Tattersall,
  Eur.\ Phys.\ J.\  C {\bf 60} (2009) 633
  [arXiv:0809.1607 [hep-ph]].

\bibitem{spheno}
  W.~Porod,
  Comput.\ Phys.\ Commun.\  {\bf 153} (2003) 275
  [arXiv:hep-ph/0301101].

\bibitem{prospino}
  W.~Beenakker, R.~Hopker and M.~Spira,
  arXiv:hep-ph/9611232.

\bibitem{Choi:2001ww}
  S.~Y.~Choi, J.~Kalinowski, G.~A.~Moortgat-Pick and P.~M.~Zerwas,
  Eur.\ Phys.\ J.\  C {\bf 22} (2001) 563
  [Addendum-ibid.\  C {\bf 23} (2002) 769]
  [arXiv:hep-ph/0108117];
  J.~Kalinowski,
  Acta Phys.\ Polon.\  B {\bf 34} (2003) 3441
  [arXiv:hep-ph/0306272];
  S.~Y.~Choi, B.~C.~Chung, J.~Kalinowski, Y.~G.~Kim and K.~Rolbiecki,
  Eur.\ Phys.\ J.\  C {\bf 46} (2006) 511
  [arXiv:hep-ph/0504122].

\bibitem{Deppisch:2009nj}
  F.~Deppisch and O.~Kittel,
  JHEP {\bf 0909} (2009) 110
  [Erratum-ibid.\  {\bf 1003} (2010) 091]
  [arXiv:0905.3088 [hep-ph]].

\end{thebibliography}
\end{document}